\documentstyle[12pt]{article}
\catcode`\@=11

\@addtoreset{equation}{section}
\def\theequation{\arabic{equation}}
\def\theequation{\thesection\arabic{equation}}

\def\a{\alpha}
\def\b{\beta}
\def\g{\gamma}

\def\e{\epsilon}

\def\NPB#1#2#3{{\it Nucl.~Phys.} {\bf{B#1}} (19#2) #3}
\def\PLB#1#2#3{{\it Phys.~Lett.} {\bf{B#1}} (19#2) #3}
\def\PRD#1#2#3{{\it Phys.~Rev.} {\bf{D#1}} (19#2) #3}

\def\JHEP#1#2#3{{\it J. High Energy Phys.} {\bf#1} (19#2) #3}
\newskip\humongous \humongous=0pt plus 1000pt minus 1000pt
\def\caja{\mathsurround=0pt}
\def\eqalign#1{\,\vcenter{\openup1\jot \caja
        \ialign{\strut \hfil$\displaystyle{##}$&$
        \displaystyle{{}##}$\hfil\crcr#1\crcr}}\,}
\newif\ifdtup

\def\@normalsize{\@setsize\normalsize{15pt}\xiipt\@xiipt
\abovedisplayskip 14pt plus3pt minus3pt%
\belowdisplayskip \abovedisplayskip
\abovedisplayshortskip  \z@ plus3pt%
\belowdisplayshortskip  7pt plus3.5pt minus0pt}
\def\small{\@setsize\small{13.6pt}\xipt\@xipt
\abovedisplayskip 13pt plus3pt minus3pt%
\belowdisplayskip \abovedisplayskip
\abovedisplayshortskip  \z@ plus3pt%
\belowdisplayshortskip  7pt plus3.5pt minus0pt
\def\@listi{\parsep 4.5pt plus 2pt minus 1pt
            \itemsep \parsep
            \topsep 9pt plus 3pt minus 3pt}}

\def\underline#1{\relax\ifmmode\@@underline#1\else
        $\@@underline{\hbox{#1}}$\relax\fi}
\@twosidetrue
\relax

\catcode`@=12

\catcode`\@=11

\def\section{\@startsection{section}{1}{\z@}{3.5ex plus 1ex minus
   .2ex}{2.3ex plus .2ex}{\large\bf}}
\def\thesection{\arabic{section}.}

\def\ps@headings{\def\@oddfoot{}\def\@evenfoot{}
\def\@oddhead{\hbox{}\hfill
        \makebox[.5\textwidth]{\raggedright\ignorespaces --\thepage{}--
        \hfill }}
\def\@evenhead{\@oddhead}
\def\subsectionmark##1{\markboth{##1}{}} }

\ps@headings

\catcode`\@=12

\relax

\def\figcap{\section*{Figure Captions\markboth
        {FIGURECAPTIONS}{FIGURECAPTIONS}}\list
        {Fig. \arabic{enumi}:\hfill}{\settowidth\labelwidth{Fig. 999:}
        \leftmargin\labelwidth
        \advance\leftmargin\labelsep\usecounter{enumi}}}
 \relax
\def\tablecap{\section*{Table Captions\markboth
        {TABLECAPTIONS}{TABLECAPTIONS}}\list
        {Table \arabic{enumi}:\hfill}{\settowidth\labelwidth{Table 999:}
        \leftmargin\labelwidth
        \advance\leftmargin\labelsep\usecounter{enumi}}}
 \relax
\def\reflist{\section*{References\markboth
        {REFLIST}{REFLIST}}\list
        {[\arabic{enumi}]\hfill}{\settowidth\labelwidth{[999]}
        \leftmargin\labelwidth
        \advance\leftmargin\labelsep\usecounter{enumi}}}
 \relax

\catcode`\@=11

\def\marginnote#1{}
\newcount\hour
\newcount\minute
\newtoks\amorpm
\hour=\time\divide\hour by60
\minute=\time{\multiply\hour by60 \global\advance\minute by-
\hour}
\edef\standardtime{{\ifnum\hour<12 \global\amorpm={am}%
    \else\global\amorpm={pm}\advance\hour by-12 \fi
    \ifnum\hour=0 \hour=12 \fi
    \number\hour:\ifnum\minute<100\fi\number\minute\the\amorpm}}
\edef\militarytime{\number\hour:\ifnum\minute<100\fi\number\minute}
\def\draftlabel#1{{\@bsphack\if@filesw {\let\thepage\relax
  \xdef\@gtempa{\write\@auxout{\string
    \newlabel{#1}{{\@currentlabel}{\thepage}}}}}\@gtempa
    \if@nobreak \ifvmode\nobreak\fi\fi\fi\@esphack}
     \gdef\@eqnlabel{#1}}
\def\@eqnlabel{}
\def\@vacuum{}
\def\draftmarginnote#1{\marginpar{\raggedright\scriptsize\tt#1}}
\def\draft{\oddsidemargin -.5truein
        \def\@oddfoot{\sl preliminary draft \hfil
        \rm\thepage\hfil\sl\today\quad\militarytime}
        \let\@evenfoot\@oddfoot \overfullrule 3pt
        \let\label=\draftlabel
        \let\marginnote=\draftmarginnote
   
\def\@eqnnum{(\theequation)\rlap{\kern\marginparsep\tt\@eqnlabel}%
\global\let\@eqnlabel\@vacuum}  }
\def\preprint{\twocolumn\sloppy\flushbottom\parindent 1em
        \leftmargini 2em\leftmarginv .5em\leftmarginvi .5em
        \oddsidemargin -.5in    \evensidemargin -.5in
        \columnsep 15mm \footheight 0pt
        \textwidth 250mmin      \topmargin  -.4in
        \headheight 12pt \topskip .4in
        \textheight 175mm
        \footskip 0pt
        
\def\@oddhead{\thepage\hfil\addtocounter{page}{1}\thepage}
        \let\@evenhead\@oddhead \def\@oddfoot{} \def\@evenfoot{}  }
\def\titlepage{\@restonecolfalse\if@twocolumn\@restonecoltrue\onecolumn
     \else \newpage \fi \thispagestyle{empty}\c@page\z@
        \def\thefootnote{\fnsymbol{footnote}} }
\def\endtitlepage{\if@restonecol\twocolumn \else  \fi
        \def\thefootnote{\arabic{footnote}}
        \setcounter{footnote}{0}}  %\c@footnote\z@ }
\catcode`@=12
\relax
\def\ps@headings{\def\@oddfoot{}\def\@evenfoot{}
\def\@oddhead{\hbox{}\hfill
        \makebox[.5\textwidth]{\raggedright\ignorespaces --\thepage{}--
        \hfill }}
\def\@evenhead{\@oddhead}
\def\subsectionmark##1{\markboth{##1}{}} }

\ps@headings

\relax
\newcommand\Appendix[1]{\def\thesection{Appendix \Alph{section}}
 \section{\label{#1}}\def\thesection{\Alph{section}}}
\def\firstpage#1#2#3#4#5#6{
\begin{document}
%%%%%%%%%%%%%%%%%%%%%%%%%%%%%%%%%%%%%%%%
%\draft
%%%%%%%%%%%%%%%%%%%%%%%%%%%%%%%%%%%%%%%%
\begin{titlepage}
\nopagebreak
\title{\begin{flushright}
        \vspace*{-1.8in}
        {\normalsize CPHT-S714.0399}\\[-9mm]
        {\normalsize LPT-ORSAY 99/12}\\[-9mm]
       %{\normalsize LPTENS-99/xx }\\[-9mm]
        {\normalsize hep-th/9906039}\\[4mm]
\end{flushright}
\vfill {#3}}
\author{\large #4 \\[1.0cm] #5}
\maketitle
\vskip -7mm     
\nopagebreak 
\begin{abstract} {\noindent #6}
\end{abstract}
\vfill
\begin{flushleft}
\rule{16.1cm}{0.2mm}\\[-3mm]
$^{\star}${\small Research supported in part by the EEC under TMR contract 
ERBFMRX-CT96-0090.}\\[-3mm] 
$^{\dagger}${\small Unit{\'e} mixte de recherche du CNRS (UMR 7644).}\\[-3mm]
$^{\ddagger}${\small Unit{\'e} mixte de recherche du CNRS (UMR 8627).}\\ 
June 1999
\end{flushleft}
\thispagestyle{empty}
\end{titlepage}}

\def\simlt{\stackrel{<}{{}_\sim}}
\def\simgt{\stackrel{>}{{}_\sim}}
\newcommand{\dal}{\raisebox{0.085cm} {\fbox{\rule{0cm}{0.07cm}\,}}}
\newcommand{\dt}{\partial_{\langle T\rangle}}
\newcommand{\dtbar}{\partial_{\langle\overline{T}\rangle}}
\newcommand{\al}{\alpha^{\prime}}
\newcommand{\mst}{M_{\scriptscriptstyle \!S}}
\newcommand{\mpl}{M_{\scriptscriptstyle \!P}}
\newcommand{\dv}{\int{\rm d}^4x\sqrt{g}}
\newcommand{\lv}{\left\langle}
\newcommand{\rv}{\right\rangle}
\newcommand{\ph}{\varphi}
\newcommand{\abar}{\overline{a}}
\newcommand{\sbar}{\,\overline{\! S}}
\newcommand{\xbar}{\,\overline{\! X}}
\newcommand{\fbar}{\,\overline{\! F}}
\newcommand{\zbar}{\overline{z}}
\newcommand{\dbar}{\,\overline{\!\partial}}
\newcommand{\tbar}{\overline{T}}
\newcommand{\taubar}{\overline{\tau}}
\newcommand{\ubar}{\overline{U}}
\newcommand{\ybar}{\overline{Y}}
\newcommand{\phb}{\overline{\varphi}}
\newcommand{\cm}{Commun.\ Math.\ Phys.~}
\newcommand{\prl}{Phys.\ Rev.\ Lett.~}
\newcommand{\pr}{Phys.\ Rev.\ D~}
\newcommand{\pl}{Phys.\ Lett.\ B~}
\newcommand{\ibar}{\overline{\imath}}
\newcommand{\jbar}{\overline{\jmath}}
\newcommand{\np}{Nucl.\ Phys.\ B~}
\newcommand{\F}{{\cal F}}
\renewcommand{\L}{{\cal L}}
\newcommand{\A}{{\cal A}}
\newcommand{\be}{\begin{equation}}
\newcommand{\ee}{\end{equation}}
\newcommand{\ba}{\begin{eqnarray}}
\newcommand{\ea}{\end{eqnarray}}
\newcommand{\dslash}{{\not\!\partial}}
\newcommand{\gsi}{\,\raisebox{-0.13cm}{$\stackrel{\textstyle >}{\textstyle\sim}$}\,}
\newcommand{\lsi}{\,\raisebox{-0.13cm}{$\stackrel{\textstyle <}{\textstyle\sim}$}\,}
\date{}
\firstpage{3118}{IC/95/34} {\large\bf Gauge couplings in four-dimensional
Type I string orbifolds} 
{I. Antoniadis$^{\,a}$ , C. Bachas$^{\,b}$ and E. Dudas$^{\,c}$}
%\\[-3mm]  
{\normalsize\sl
$^a$ Centre de Physique Th{\'e}orique$^\dagger$  Ecole Polytechnique,
{}F-91128 Palaiseau, France\\[-3mm]
\normalsize\sl$^b$ Laboratoire de Physique Th{\'e}orique, ENS, F-75231 Paris, France\\[-3mm]
\normalsize\sl $^c$  LPT$^\ddagger$, B{\^a}t. 210, Univ. Paris-Sud, F-91405 Orsay,
France} 
{We compute threshold effects to gauge couplings in four-dimensional $Z_N$
orientifold models of type I strings with ${\cal N}=2$ and ${\cal N}=1$
supersymmetry, and study their dependence on the geometric moduli. 
We also compute the tree-level (disk) couplings of the open sector
gauge fields to the twisted closed string moduli of the orbifold in
various models and study their effects and that of the one-loop
threshold corrections
on gauge coupling unification. We interpret the results from the
(supergravity) effective theory point of view and comment on the conjectured 
heterotic-type I duality.}
%%%%%%%%%%%%%%%%%%%%%%%%%%%%%%%%%%%%%%%%%%%%%%%%%%%%%%%%%%%%%%%%%%%%%%%%%%%%%%%%%%%%%%%%
\section{Introduction}

The purpose of this paper is to study threshold corrections to
gauge couplings in certain ${\cal N}=2$ and ${\cal N}=1$ type I
orientifolds \cite{Sa}-\cite{ADS}. Such vacua are of obvious
phenomenological interest, but
for a long time  they had  received  much less attention than their
weakly-coupled heterotic counterparts. As has been, however,  recognized
recently,
type I vacua  offer
some added flexibility for model building, and can exhibit several novel
interesting features. In particular, since gauge interactions  are localized
on  D-branes the tree-level relations between  gauge
couplings and the string, compactification  and Planck scales are not
universal \cite{AS,W}, making it easier to consider scenarios in which these
scales are hierarchically-different \cite{A}-\cite{AP}. The question of
whether such scenaria  can be reconciled with the apparent unification
of the observed low-energy couplings was  the main motivation for this work.

Type I vacua are furthermore conjectured to have dual heterotic
descriptions in appropriate circumstances \cite{PW}-\cite{ABFPT}.
Thus, another motivation for the present work
has been to further elucidate this duality in the
most interesting ${\cal N}=1$ context \cite{ABPSS,K}, by a detailed
comparison of threshold corrections on the two sides,
even if they cannot be simultaneously weakly coupled in ten dimensions.
Moreover, the knowledge of the moduli dependence of gauge couplings
is required for any phenomenological application of type I string theory,
and is intimately related to the problems of supersymmetry breaking and
possible dynamical determination of the moduli vacuum expectation values (VEVs).

Threshold corrections to gauge couplings in heterotic vacua have been studied extensively
in the past and general results have been obtained for orbifold and smooth
manifold compactifications \cite{Ka,DKL}. Their comparison to the
effective field theory is particularly instructive and reveals the
existence of a universal non-holomorphic correction associated to the so
called Green-Schwarz term \cite{DFKZ,KL2}. On the type I side on the other hand,
previous study was focused on the ${\cal N}=2$ orientifold based on the
$T^4/Z^2$ orbifold \cite{BF,apt}.

In this work, we first generalize the studies to other
${\cal N}=2$ orientifolds obtained by
toroidal compactifications of six-dimensional (6d) vacua and compute the
general one-loop dependence of gauge couplings on the geometric moduli 
$T\sim R_1R_2$ and $U\sim R_1/R_2$ of the two-torus with radii $R_{1,2}$. For vanishing
Wilson lines, the corrections are proportional to the beta function coefficients and
therefore they amount changing the unification scale from the string to the
compactification size. The latter can be much larger than the string scale in a weakly
coupled theory, in which case it is more appropriate to identify it as the winding
scale of the T-dual theory. 
This result allows two possibilities for unification.
(i) When the two radii are different $R_1 > R_2$, there is a 
linear correction $R_1/R_2$ \cite{B} that can be interpreted as a power-law evolution
\cite{TV,DDG} between
$1/R_1$ and $1/R_2$, leading to unification at $1/R_2$ \cite{DDG}. (ii) If the two
radii are approximately equal $R_1 \simeq R_2$, there is only a logarithmic
correction which could possibly accomodate a ``conventional"
unification scenario at energies much 
larger than the string scale \cite{B,AB}.
Threshold corrections can alternatively be interpreted
as tree-level dependence of bulk fields on the transverse space \cite{AB}.
These corrections are non-universal (group dependent) if
tree-level couplings to bulk fields are different
for the various gauge group factors.
In fact, there is a tree-level (disk) dependence
of gauge couplings on the twisted Neveu-Schwarz (NS) moduli associated to
twists different than $Z_2$'s. This dependence is already present at the
level of six-dimensional theory, where these moduli are part of tensor
multiplets \cite{S,DM}. Although in some particular examples this new (tree-level)
contribution is miraculously proportional to the (one-loop) beta-function
coefficients, in which case the unification scale is arbitrary, in the generic case
the couplings become free parameters and the unification is lost.

These results are generalized easily in the case of 4d compactifications
with ${\cal N}=1$ supersymmetry. The geometric moduli dependence of the
one-loop threshold corrections receives contributions from the ${\cal N}=2$
sectors only, controled by the ${\cal N}=2$ beta-functions, in analogy with the
heterotic case. On the other hand,
${\cal N}=1$ sectors provide moduli independent contributions that can be interpreted
as one-loop running up to the string scale. As a result, if the string scale is low
the only way to achieve unification is through the ${\cal N}=2$ beta-functions.
However, as in  ${\cal N}=2$ compactifications, there is a non-universal
tree-level dependence of the gauge couplings on the twisted NS-NS moduli for twists
different than $Z_2$'s \cite{IRU}, which apriori can destroy unification. In this case
though, a new phenomenon appears due to the existence of anomalous $U(1)$'s. At the
points of maximal gauge symmetry, the VEVs of the twisted moduli are fixed and 
unification is recovered up to one-loop level.
 
We also compare our results to the effective supergravity. Unlike the
heterotic case, we find that compatibility requires the existence of
several {\it non-universal} (gauge group dependent) Green-Schwarz terms
associated to the twisted NS-NS moduli \footnote{This result was also obtained
recently in ref. \cite{IRU2}, while our work was in print.}
which now belong into several linear
multiplets. In the ${\cal N}=1$ orientifold examples the vanishing of the
anomalous $U(1)$'s D-terms at the points of maximal gauge symmetry, imply a
vanishing VEV for the twisted moduli coupled to gauge fields, in the {\it string}
(linear multiplet) basis. This implies in particular for the $Z_{\rm odd}$ orbifolds
that physical gauge couplings have no dependence on geometric moduli up to one-loop
level and unification arises at the type I string scale \footnote{This result seems to
contradict the recent claim of ref. \cite{I}.}. Moreover, this result raises
doubts on the validity of the conjectured heterotic -- type I dual pairs with 
${\cal N}=1$ supersymmetry.

The paper is organized as follows. In section 2, we recall open string
propagation in constant magnetic fields, that provides the method we use for
computing the corrections to gauge couplings. In section 3, we derive the
moduli dependence to gauge couplings of generic 4d ${\cal N}=2$
orientifolds obtained by toroidal compactification from six dimensions. In
section 4, we study in detail gauge coupling in the context of the 4d ${\cal N}=1$
$Z_{\rm odd}$ ($Z_3$ and $Z_7$) orientifolds, while in section 5, we study
the ${\cal N}=1$ $Z_6^\prime$ and
we obtain general expressions valid for any orientifold. In section 6,
we compute threshold corrections in models with spontaneous
${\cal N}=4\to{\cal N}=2$ supersymmetry breaking. In section 7, we
compare our results with the effective field theory and comment on
heterotic -- type I duality. In section 8, we discuss anomalous $U(1)$'s and
implications to gauge coupling unification. Our conclusions are presented in section 9.

%%%%%%%%%%%%%%%%%%%%%%%%%%%%%%%%%%%%%%%%%%%%%%%%%%%%%%%%%%%%%%%%%%%%%%%%%%%%%%%%%%%%%%%
\section{Background-field method}

The one-loop diagrams of type-I string theory are the  torus ${\cal T}$, the
Klein bottle ${\cal K}$, the annulus ${\cal A}$,  and the M{\"o}bius
strip ${\cal M}$ . The
diagrams with boundaries (${\cal A}$ and ${\cal M}$)  describe  in the direct
channel the propagation of the
open string degrees of freedom, including 
gauge bosons  and  charged matter fields.  The other two diagrams
(${\cal T}$
and ${\cal K}$) have no boundaries coupling to Chan-Paton charges -- they 
describe the propagation in the loop 
of closed-string degrees of freedom  related to the gravitational
sector of the theory. Since the ${\cal T}$ and ${\cal K}$  diagrams cannot couple to
external open-string states, they do not contribute to the
renormalization of  couplings 
in  the open sector of the theory. These diagrams   will not be
of interest to us in the present work.

 We will here focus our attention to  four-dimensional  orientifolds
obtained by orbifolding the six real (three complex) internal
coordinates by the  twist
$\theta = (e^{2 i\pi v_1},e^{2 i\pi v_2} ,e^{2 i\pi v_3} )$, where ${\bf v} \equiv
(v_1,v_2,v_3)$ is called the twist vector and where for a $Z_N$ orbifold 
$\theta^N = 1$. The two  one-loop amplitudes of interest  can be
written generically as \vskip 0.1cm
\be\eqalign{
\ \ \ \ \ {\cal A} \ = &\ -{1 \over 2N}  \sum_{k=0}^{N-1} \int_0^{\infty} 
{dt \over t}\  
Str  \left( \theta^k \; { q}^{(p^{\mu}p_{\mu}+m^2)/2}\right)\; 
 \equiv \;-{1 \over 2N} \sum_{k=0}^{N-1}
\int_0^\infty {dt \over t} {\cal A}^{(k)}(q) \ , \cr
\ \ \ \ \ {\cal M} \ = &\  -{1 \over 2N}   \sum_{k=0}^{N-1} \int_0^{\infty} 
{dt \over t} \; Str
 \left( \Omega  \theta^k \; { q}^{(p^{\mu}p_{\mu}+m^2)/2}\right)\; \equiv
\;
-{1 \over 2N} \sum_{k=0}^{N-1}
\int_0^\infty  {dt \over t}  {\cal M}^{(k)}(-q) \ ,\cr }  \label{u1}
\ee
\vskip 0.10cm
\noindent  where $m^2$ is the mass squared operator in four
dimensions,
the Regge slope $\alpha^\prime=1/2$, 
 the modular parameter of the
`doubling torus' (from which ${\cal A}$ and ${\cal M}$ are obtained by
a $Z_2$ identification)  is given by 
\be
\tau ={it \over 2} \ \ \ \ {\rm for} \ \ {\cal A}   \ ,  \ \ {\rm and} \ \ \ 
\ \tau = {it \over 2} + {1 \over 2} \ \ \ \ 
  {\rm for}\ \ {\cal M} \ ,   \label{u01}
\ee
and finally  $q =e^{-\pi t}$.
In the M{\"o}bius amplitude  
$\Omega$ is the world-sheet
involution operator which exchanges  left and right moving excitations of the closed string
and acts  as a phase on the open string oscillators~:  $\Omega \alpha_m = \pm e^{i \pi m}
\alpha_m$, with the upper plus sign for the NN coordinates  and the lower minus sign for the
DD coordinates  (N=Neumann, D=Dirichlet). The supertrace stands for a
sum  over all open-string states -- this includes the sum over
Chan-Paton states of the
two endpoints  and the  integration over four-momenta, 
\be
Str = ( \sum_{bos} - \sum_{ferm} )\;  \int {d^4 p \over (2 \pi)^4 }   \ . \label{u2}
\ee 
The action of a twist element $\theta^k$ on the $n$ possible Chan-Paton states is
realized by $n \times n$
matrices $\g^k $. In the simplest cases we have   $n=32$,  corresponding to the
32 D9-branes of the ten-dimensional theory. More generaly however the
four-dimensional  orientifold may also contain D5-branes,
corresponding to additional
Chan-Paton states. Conversely $n$ and the rank of the gauge group can
be reduced by turning on $B_{\mu \nu}$ backgrounds in the internal
manifold \cite{BPS}. 
Tadpole consistency conditions \cite{PC} severely constrain   the Chan-Paton
matrix $\g$, and hence also  the gauge group and the charged matter content of the
corresponding vacuum.

  We will compute the one-loop corrections to the gauge couplings
using  the background-field method \cite{BP,BF}. To this end we turn
on a  magnetic field  in the (for example) $x^1$ direction
\be
F_{23} = B Q \ , \label{u3}
\ee
where  $x^0 \cdots x^3$ are the uncompactified spacetime dimensions,
and $Q$ is an appropriately normalized   generator of the gauge group.  The effect 
of the magnetic field on the open-string spectrum \cite{ACNY} 
is to  shift  the oscillator modes
of the complex $X^2+iX^3$ string coordinate by an amount $\e $, where
\be
\pi \e = \arctan (\pi q_L B) +  \arctan (\pi q_R B) \ , \label{u4}
\ee
and $q_{L(R)}$ is the eigenvalue of the gauge-group generator $Q$ acting on the
Chan-Paton states at the left(right) endpoint of the open string. 
For notational simplicity, the dependence of $\e$ on the Chan-Paton
states  will be left implicit in the sequel.
The torus and Klein-bottle amplitudes are not
affected by the magnetic field, while the annulus and the M{\"o}bius
strip  are obtained by making in (\ref{u1}) the replacements
\cite{BP,ACNY}
\ba
p^{\mu} p_{\mu} & \rightarrow & -(p_0)^2 + (p_1)^2 + (2n+1) \e + 2 \e \Sigma_{23} \ , 
\nonumber \\
Str & \rightarrow & ( \sum_{bos} - \sum_{ferm} )
 {(q_L+q_R) B \over 2 \pi} \sum_n \int {d^2 p \over (2 \pi)^2 }
 \ , \label{u5} 
\ea
where $\Sigma_{23}$ is the spin operator in the $(23)$ direction,
 the integer $n=0,1, \cdots$
labels the Landau levels, and  $(q_L+q_R) B/ 2 \pi$ is the
degeneracy of Landau levels per unit area. 

%%%%%%%%%%%%%%%%%%%%%%%%%%%%%%%%%%%%%%%%

The full one-loop vacuum energy has the weak-field expansion
\be
\eqalign{
\Lambda (B&) = {1 \over 2} \Bigl( {\cal T}+{\cal K} +  {\cal A}(B) +  
 {\cal M}(B) \Bigr) \cr
\equiv & \ \Lambda_0 + {1 \over 2} \left({B\over 2\pi}\right)^2
 \Lambda_2 + {1 \over 24}\left({B\over 2\pi}\right)^4 \Lambda_4 + 
\cdots \ .
\cr } \label{u6}
\ee
For supersymmetric compactifications the one-loop cosmological
constant vanishes, $\Lambda_0=0$.
The term quadratic in the background  field contains  the 
one-loop threshold corrections.  More precisely, if 
we choose $Q$ to be an appropriately  normalized generator
\footnote{The factor groups in the examples of interest
will have a natural embedding into $SO(32)$. Thus will consider  the $F_{\mu\nu}$'s
as  matrices in  the fundamental
representation of $SO(32)$, with  generators
normalized  so that ${\rm tr} Q^2   = 1$. We will write the 
 Yang-Mills lagrangians as   $ {\cal L}_{YM}= {\rm tr}_a
  F_{\mu\nu}F^{\mu\nu}/ 4g^2_a$ -- with these conventions 
the tree-level gauge couplings of the various factors $G_a$ are all  equal.} 
inside the $a$th  factor of the gauge group, then
the one-loop corrected gauge coupling for  this factor reads
\be 
{4\pi^2 \over g_a^2}\Biggl\vert_{\rm \ one \ loop}\ 
 = \ {4\pi^2 \over g_{a}^2} \Biggl\vert_{\rm \ tree} \  + \
 \Lambda_{2,a}   \ \ .  \label{u7}
\ee
The loop correction has the structure of an integral
\be
\Lambda_{2,a} =  \int_0^{\infty} {dt \over 4t}\;
{\cal B}_a (t) \ \ , \label{u07}
\ee
with the upper and lower limits corresponding, respectively, to the ultraviolet
and the infrared regions in the open channel. 
As explained  in references 
\cite{BF},\cite{B} in the context of the $Z_2$ orientifold \cite{BS},
the  integral must converge  in the ultraviolet limit  
if all the tadpoles have been  cancelled globally,  and provided that the
background field has no
component along an  anomalous $U(1)$ factor. The potential  infrared
divergences, on the other hand, are due to massless charged particles
circulating in the loop, so that 
 \be
 {\rm lim}_{t\to\infty}\ {\cal B}_a (t) = b_a 
 \ee
is the $\beta$-function coefficient of the effective field theory
at energies much lower than  the last massive threshold. 

The quartic term in the expansion (\ref{u6}) is quadratically divergent
due to the on-shell exchange of  massless closed-string modes
\cite{BF}, \cite{BK}. These include  the dilaton,  the 
graviton and, in the cases of interest to us here,  the  twisted NS-NS
moduli fields $m_k$ of the orbifold. 
These twisted moduli  have non-universal couplings to the gauge fields of the
gauge group factor $G_a$, 
\be
{\cal L}_{YM} = \Bigl( {1\over  4g^2_a} + \sum_{k=1}^{[{N-1 \over 2}]}
{s_{ak}\over 16\pi^2}\; m_k \Bigr)\;  {\rm tr}_a
 F_{\mu\nu}F^{\mu\nu}\ , \label{ym}
\ee
as conjectured, using anomaly-cancellation arguments and supersymmetry, 
in ref. \cite{S,DM,IRU}. In eq. (\ref{ym}), the sum over $k$ goes up to
the integer part of $(N-1)/2$ which counts the number of independent twisted 
sectors of the orbifold.
Note that the supersymmetric partners of the $m_k$ are  
RR axions, dual to antisymmetric two-index tensors (NS=Neveu-Schwarz, R=Ramond).
By analyzing  the divergences  of $\Lambda _4$  we will 
calculate the coefficients $s_k$ explicitly, and 
confirm  the conjecture of \cite{IRU}. This is important 
for discussions of unification, since such twisted modes
could give rise to non-universal shifts of the gauge coupling constants at
tree level.

In order to analyze the  divergences of $\Lambda_4$ we must reexpress
the amplitude as an integral over
the modulus $l$ in the transverse (closed-string) channel. This is 
 related to the direct 
channel modulus as follows,
\ba
  l = {1 \over t} \ \ {\rm for}\ {\cal A} ,  \quad l = {1 \over 4t} 
\ \ {\rm for}\ {\cal M}\ . \label{u8}
\ea
The elliptic functions in the integrands can be reexpressed in terms
of $l$ by using the  modular transformations 
\be
\tau = {i t \over 2} \rightarrow -{1 \over \tau} = 2 i l \  \label{u9}
\ee
for ${\cal A}$, and 
\be
\tau = {i t \over 2} +{1 \over 2} \rightarrow -{1 \over \tau} \rightarrow
- {1 \over \tau}+2 \rightarrow ({1 \over \tau}-2)^{-1} = 2 i l - {1 \over 2} \  
\label{u10}
\ee
for ${\cal M}$. After this change of variable the quartic term will
take the form
\be
\Lambda_4  = -{1 \over 4 N} \sum_{k=0}^{N-1}\;\;\;
 \int_0^\infty  {dl \over l} \;  \left\{
 {\cal A}^{(k)}_4 ({\tilde q})+ {\cal M}^{(k)}_4(-{\tilde q}) \right\} \ , \label{u010}
\ee
where ${\tilde q}= e^{-4\pi l}$,  and ${\cal A}^{(k)}_4$ and ${\cal
M}^{(k)}_4$
are the coefficients in the Taylor expansion of the corresponding
integrands  at  quartic order
in $B/2\pi$. These grow linearly at $l\to\infty$, corresponding to a
quadratic infrared divergence in the closed-string channel. 
The divergence  in the untwisted ($k=0$)
sector comes from the exchange of a graviton and dilaton and has been
analyzed in \cite{BF,BK}. For even $N$
the $k= N/2$ sector has  no quadratic  divergences, consistently
with the fact that  $Z_2$ twist fields do not couple to the
Yang-Mills action. \footnote{This is obvious in six dimensions, where
the $Z_2$ twist fields belong to  hypermultiplets that do not couple to the vectors.}
For the remaining  sectors ($k=1,\cdots 
[{N-1 \over 2}]$) we will show that
\be
{\cal A}^{(k)}_4 + {\cal M}^{(k)}_4  =   {3 N \pi } \;  l \; s_k^2 +
\cdots 
\ , \label{u011}
\ee
where the dots stand for  exponentially-suppressed terms  and the precise
values of the coefficients $s_k$ depend on the model. 
This result is consistent with our interpretation of the corresponding 
divergence in $\Lambda_4$, 
as coming from the  exchange of an on-shell twist field coupling  to the
background $F_{\mu\nu}$  through equation (\ref{ym}). Notice that the
 closed-string propagator for a canonically normalized scalar is
\be
\Delta_{closed} = {\pi \over 2} \int_0^{\infty} \ dl \ e^{-{\pi l \over 2}
(p^{\mu}p_{\mu} + M_{closed}^2)} \ , \label{f120}
\ee
with $l$ the  modulus of the cylinder. The divergence of an
on-shell propagator can thus be written formally as ${\pi\over
2}\int^\infty dl$.

%%%%%%%%%%%%%%%%%%%%%%%%%%%%%%%%%%%%%%%%%%%%%%%%%%%%%%%%%%%%%%%%%%%%%%%%%%%%%%%%%%%%%%%
\section{ ${\cal N}=2$ supersymmetry:  $K3\times T^2$ orientifolds}

   We begin  our discussion  with  six-dimensional $Z_N$  models 
\cite{BS,GJ,DP}, compactified further down to four dimensions on a
two-torus. The twist vector for these models is of  the form
 ${\bf v}=(1/N,-1/N,0)$. Assuming no
antisymmetric tensor backgrounds, tadpole cancellation 
 requires the presence of 32 D9
branes,  and  for even $N$  of one set of 32
D5 branes. It also  fixes the matrices  $\gamma_9$ and $\gamma_{\Omega,9}$
which represent the action of the orbifold twist $\theta$ and the
orientation reversal $\Omega$  on
Chan-Paton states in the D9 sector, as well as the corresponding ones
in the D5-brane sector.\footnote{ 
Consistently with the group property 
the action of $\theta^k$ and 
of $\Omega\theta^k$ 
can be  represented by the product matrices 
$\gamma^k$ and  $\gamma^k \gamma_{\Omega}$ respectively
.} 
 The particular case of the $Z_2$ orientifold, $N=2$,  is the
one  analyzed previously in reference \cite{BF}. In this section 
 we will extend the
analysis  to the 
other models of type A (using the language of \cite{GJ}), namely the $Z_3,
Z_4$ and $Z_6$ orbifolds. 

   We will consider a background field living on the D9 branes of
these models. To simplify the formulae we will further restrict our
attention to the case where all D5-branes are located together
at a fixed point of the
orbifold, and there are no (99) Wilson lines. 
The amplitudes of interest are the (99)  annulus and
M{\"o}bius diagrams with insertion of a non-trivial twist $\theta^k$ ($k\not=0$)
and,  in the presence  of D5 branes,  the (95) annulus with or without
insertion of a twist. The (55) diagrams do not couple to the
background field,
 while the untwisted (99) diagrams have effectively ${\cal N}=4$
supersymmetries and thus do not contribute to the renormalization of gauge
couplings. In the absence of a magnetic field the relevant amplitudes
 are (for $k\not=0$)  \vskip 0.02cm
\ba
\eqalign{
{\cal A}_{99}^{(k)} \!&=\! - {\rm tr} ( {\g}_9^{k}\otimes  {\g}_9^{k})\;
 {\Gamma^{(2)}(t)\over 4\pi^4t^2}  \sum_{\a,\b=0,1/2}  {1\over 2} \eta_{\a,\b} \
\frac{\vartheta^2[{\a \atop \b}]}{\eta^6}
 \times \! (2\sin {\pi k\over N})^2
\  \frac{\vartheta[{\a \atop {\b + k/N}}] \vartheta[{\a \atop {\b - k/N}}]}
{\vartheta[{1/2 \atop {1/2 +k/N}}] \vartheta[{1/2 \atop {1/2 -k/N}}]} \,  \
  \ , \cr
{\cal M}_{99}^{(k)} &=\  
 {\rm tr}({\g}_9^{2k}) \; {\Gamma^{(2)}(t)\over 4\pi^4t^2} 
\sum_{\a,\b=0,1/2}  {1\over 2} \eta_{\a,\b} \
\frac{\vartheta^2[{\a \atop \b}]}{\eta^6}
 \times 
 (2\sin {\pi k\over N})^2
\  \frac{\vartheta[{\a \atop {\b + k/N}}] \vartheta[{\a \atop {\b - k/N}}]}
{\vartheta[{1/2 \atop {1/2 +k/N}}] \vartheta[{1/2 \atop {1/2 -k/N}}]} \,  \
 \ ,
\cr } \label{f1}
\ea 
\vskip 0.02cm \noindent  and if
  the model has D5-branes (N  even, all $k$) \vskip 0.02cm
\ba
{\cal A}_{95}^{(k)} = 2\;
 {\rm tr} ( {\g_9}^{k}\otimes {\g_5}^{k})\;
  {\Gamma^{(2)}(t)\over 4\pi^4t^2} 
 \sum_{\a,\b=0,1/2} {1\over 2}  \eta_{\a,\b} \
\frac{\vartheta^2[{\a \atop \b}]}{\eta^6} \
  \times 
\  \frac{\vartheta[{\a+1/2 \atop {\b + k/N}}]
 \vartheta[{\a +1/2\atop {\b - k/N}}]}
{\vartheta[{0 \atop {1/2 +k/N}}] \vartheta[{0 \atop {1/2 -k/N}}]} \,  \
 \ . \label{f2}
\ea
\vskip 0.02cm 
\noindent  In these expressions $\eta_{\a, \b}=(-1)^{2(\a+\b+2 \a \b)}$ are the
usual phases 
depending on the spin structures $(\a,\b)$ and  specifying the GSO
projection .
The definitions of the Jacobi functions ${\vartheta[{\a \atop \b}]}$
and of the Dedekind function $\eta$ are given in appendix A, while  their
  argument (\ref{u01})
is  here left implicit. The factor
of 2 in front of the (95) diagram  counts the two orientations of the
open string, and $\Gamma^{(2)}(t)$
is the lattice sum over
momenta along the untwisted two-torus.
 Finally, the trace in the
annulus amplitudes is over the tensor product of Chan-Paton states  for
the left and the right endpoints of the open string, 
while in  the M{\"o}bius amplitude the left and right Chan-Paton
charges  are equal and  we
have used the identity 
 ${\rm tr} {\g}^{2k}=  {\rm tr} ({\g}_{\Omega\theta^k}^{-1}
 {\g}_{\Omega\theta^k}^T)$ (see for instance 
\cite{GJ}, \cite{AFIV}).

%%%%%%%%%%%%%%%%%%%%%%%%%%%%%%%%%%%%%%%%%%%%%%%%%%%%%%%%%%%%%%%%%

The above vacuum amplitudes vanish of course by virtue  of the space-time
supersymmetry. Turning on the background magnetic field modifies these
expressions
 as follows,
\be\eqalign{
{\cal A}_{99}^{(k)}(B) = -
 iB {  \Gamma^{(2)}(t) \over 4 \pi^3 \; t}\; 
 \sum_{\a,\b=0,1/2}{1\over 2} \eta_{\a,\b} \frac{\vartheta[{\a \atop \b }]}{\eta^3}&\ 
 {\rm tr}\left( (Q\g_9^k\otimes \g_9^k +
\g_9^k\otimes Q \g_9^k)
 \frac{\vartheta[{\a \atop \b}]({i \e t\over 2}) }{\vartheta[{1/2 \atop 1/2
     }]({i \e t\over 2})}
\right) \cr\times & 
 (2\sin {\pi k\over N} )^2
\frac{
\vartheta[{\a \atop {\b + k/N}}] \vartheta[{\a \atop {\b - k/N}}]}
 {  \vartheta[{1/2 \atop 1/2 +k/N}]
\vartheta[{1/2 \atop 1/2 -k/N}]}\ ,  \cr}
\ee
\be
\eqalign{
\ \ \ \ {\cal M}_{99}^{(k)}(B) = 
 iB {  \Gamma^{(2)}(t) \over 2 \pi^3 \; t}\; 
 \sum_{\a,\b=0,1/2}{1\over 2} \eta_{\a,\b} \frac{\vartheta[{\a \atop \b }]}{\eta^3} \ 
 {\rm tr}& \left(  Q\g_9^{2k} 
\frac{\vartheta[{\a \atop \b}] ({i \e t\over 2}) }
{\vartheta[{1/2 \atop 1/2}]({i \e t\over 2})}
\right)
 \cr \times   &
 (2\sin{\pi k\over N} )^2
\frac{
\vartheta[{\a \atop {\b + k/N}}] \vartheta[{\a \atop {\b - k/N}}]}
 {  \vartheta[{1/2 \atop 1/2 +k/N}]
\vartheta[{1/2 \atop 1/2 -k/N}] }\ , 
 \cr} 
\ee
and
\be
\eqalign{
{\cal A}_{95}^{(k)}(B) = 
 iB {  \Gamma^{(2)}(t) \over 2 \pi^3 \; t}\; 
 \sum_{\a,\b=0,1/2} {1\over 2} \eta_{\a,\b} \frac{\vartheta[{\a \atop \b }]}{\eta^3} \ 
 {\rm tr} & \left(  (Q\g_9^{k}\otimes \g_5^k) 
\frac{\vartheta[{\a \atop \b}] ({i \e t\over 2})  }{\vartheta[{1/2
    \atop 1/2}] ({i \e t\over 2}) }
\right) \cr &
\times  
\frac{\vartheta[{\a +1/2 \atop {\b + k/N}}] \vartheta[{\a +1/2 \atop {\b - k/N}}]
}
 { \vartheta[{0 \atop 1/2 +k/N}]
\vartheta[{0 \atop 1/2 -k/N}] }\ . \cr}
\ee
\vskip 0.3cm   As previously, the $\tau$ argument of the Jacobi functions
$\vartheta[{\a \atop \b}](z\vert\tau)$ is implicit, while
 the argument $z$ is only shown when it is non-zero.
Note that the shift
 $\epsilon$ of the oscillator frequencies 
depends on the charges of the left and the right
string endpoint -- the corresponding Jacobi functions have therefore been left
inside the Chan-Paton trace. More explicitly  in view of the
definition (\ref{u4})
we have
\be
\pi\epsilon  = \cases{&
${\rm arctan}(\pi BQ) \otimes 1 + 1\otimes {\rm
  arctan}(\pi BQ) $ \ \ \   in\  ${\cal A}_{99}$ , 
\cr
&  $2\; {\rm arctan}(\pi BQ) $ \ \ \ \ \ \ \ \ \ \ \ \ 
\ \ \ \ \ \ \ \ \ \ \ \ \ \ \ \ \ \ \ \ \   in\  ${\cal M}_{99}$,
\cr 
&  $ {\rm arctan}(\pi BQ) $\ \ \ \ \ \ \ \ \ 
\ \ \ \ \ \ \ \ \ \ \ \ \ \ \ \ \ \ \ \ \ \ \ \ \ \ \   in\  ${\cal A}_{95}$. 
\cr} 
\ee      
 The lattice sum and the orbifold
partition function, on the other hand, are independent of the
Chan-Paton states since we have assumed vanishing  Wilson lines.

%%%%%%%%%%%%%%%%%%%%%%%%%%%%%%%%%%%%%%%%%%%%%%%%%%%%

   In order to compute threshold corrections, we must expand the above
 formulae to quadratic order in the background field. We need
the following Taylor expansions  
\be
\eqalign{
\e \; \simeq\; &(q_L+q_R)B +o(B^3) \ \ ,\cr
\vartheta_1(z) \; &\simeq\;
   2\pi\eta^3\; z\; +\; o(z^3)\ \ , \cr
\vartheta_{a}(z)\;  &\simeq\; \vartheta_{a} +
{z^2\over 2} \vartheta_{a}^{\prime\prime} \; + \; o(z^4)
\ \ {\rm for}\ \ a =2,3,4\ \ . 
\cr} 
\ee
The integrands simplify enormously if one uses the modular identities
(\ref{a2}) of appendix A which reduce the  entire string-oscillator
sum to a  number. As explained in \cite{BF},\cite{DL},\cite{BK}
this is a consequence of ${\cal N}=2$ supersymmetry: only short BPS
multiplets can contribute corrections to the gauge couplings, and all
open-string excitations are non-BPS. After some straightforward
algebra the final result for the one-loop corrections (\ref{u07})
takes the form
\be
{\cal B}_a(t) =   b_a\times  \Gamma^{(2)}_{\rm reg}(t) \label{f777}
\ee
with 
\be
b_a = 
-{1 \over N}  \sum_{k=0}^{N-1} \left\{  4 \sin^2 {\pi k\over N}
\left [ {\rm tr} (Q_a^2 \g_9^k) {\rm tr} \g_9^k- 2\; {\rm tr}( Q_a^2 \g_9^{2k})
\right] -
 {\rm tr} (Q_a^2 \g_9^k)
{\rm tr} \g_5^k \right\} .  \label{f77}
\ee
We have here used the obvious identity  ${\rm tr}D\otimes E = {\rm tr}D{\rm
 tr}E$, and the fact that if $Q$ does not have a component along an
   anomalous U(1), all  the traces involving  an odd power of $Q$ are zero.  
Note also  that  $Q$ must commute  with  $\gamma_9$  or else the
corresponding gauge field  would not have survived the orbifold projection.

%%%%%%%%%%%%%%%%%%%%%%%%%%%%%%%

    The lattice sum in equation (\ref{f777}) has been  regularized by the 
`Poisson-resummation  prescription' \cite{BK,B}. Let us introduce the usual
geometric modulus of the two-torus (which spans the dimensions 4 and 5),  
\be 
U = {G_{45}+i {\sqrt G} \over G_{55}}  \ , \  \label{f14} 
\ee 
normalized so that ${\rm Im U}=R_4/R_5$ and ${\sqrt G} =R_4R_5$ 
on a rectangular torus. The regularized  Kaluza-Klein  sum is 
\be
\Gamma^{(2)}_{\rm reg}(t) = \sum_{n_4,n_5} e^{-\pi t\vert
n_4+n_5U\vert^2/({\sqrt G}{\rm Im U})} - {\pi {\sqrt G }\over t}\ .
 \label{sub}
\ee
The subtraction term corresponds to the propagation of massless
closed-string states in the transverse channel -- it vanishes by
global tadpole cancellation after adding all diagrams and imposing  a
homogeneous cutoff in the transverse proper time $l$.
\footnote{This amounts to different ultraviolet
cutoffs in  the direct channel for the annulus and the M{\"o}bius
diagrams, see  the relations (\ref{u8}).} We will explain how this
happens in more detail in a minute. In the limit  $t\to\infty$ we find
$\Gamma^{(2)}_{\rm reg}\to 1$, so that 
$b_a$ must  the $\beta$-function coefficient
of the four-dimensional theory.

We have checked that in all the models of \cite{GJ}
expression (\ref{f77}) coincides with the standard  expression for the ${\cal N}=2$
$\beta$-functions,
\be
b_a = 2\sum_r T_a(r) - 2 T_a(G)\ ,
\ee
where the hypermultiplets transform in the representations $r$ of the gauge group $G_a$. 
The gauge groups, charged hypermultiplets, and the Chan-Paton matrix
$\gamma_9$ for  these models are summarized  for convenience in table 1. 
Following the conventions of \cite{GJ} we use a complex basis in which
$\g_9$ is a diagonal matrix whose eigenvalues are phases occuring in
complex-conjugate pairs. For instance for the $Z_3$ model  
\be
\g_9 = {\rm diag}\left(\ e^{2\pi i/3} ({\rm 8 \ times})\ \  e^{-2\pi i/3} ({\rm 8
  \ times})\ \ 
 1 ({\rm 16 \ times}) \right)\ .
\ee
In a self-explanatory notation we will write
$\gamma_9= \left(e^{2\pi i\over 3}{\rm I}_8;\; {\rm I}_8 \right)$ 
to denote this matrix. Note also that the twist matrix for the
D5-branes can be chosen such that $\gamma_5= e^{2\pi i m \over N}
\gamma_9$ for any odd integer $m$ \cite{GJ}.

%%%%%%%%%%%%%%%%%%%%%%%%%%%%%%%%%%%%%%%%%%%

\vskip 0.9cm

\begin{tabular}{|l|c|c|c|}   \hline
 {Model }& {$\gamma_9$} & {99 Gauge Group}  &
{Charged Hypermultiplets}\\
 \hline\hline
$Z_2$&
$ (i {\rm I}_{16})$&U(16)&2$\times${\bf 120}; 16$\times${\bf 16} \\ \hline
$Z_3$&
$(e^{2\pi i\over 3}{\rm I}_{8},\; {\rm I}_{8})$&U(8)$\times$SO(16)&({\bf 28,1}); ({\bf 8,16}) \\ \hline
$Z_4$&
$(e^{\pi i\over 4}{\rm I}_{8},\; e^{3\pi i\over 4}{\rm
I}_{8})$&U(8)$\times$U(8)&({\bf 28,1});
 ({\bf 1,28}); ({\bf 8,8})\\
{}&{}&{}&
8$\times$({\bf 8,1}); 8$\times$({\bf 1,8}) \\ \hline
$Z_6$&
$(e^{\pi i\over 6}{\rm I}_{4},\; e^{5\pi i\over 6}{\rm I}_{4},\; i{\rm I}_{8} )$
&${\rm U(4)}^2$$\times$U(8)& $({\bf 6,1,1});$({\bf 1,6,1});$({\bf 4,1,8});$({\bf 1,4,8})\\ 
{}&{}&{}&
4$\times$({\bf 4,1,1}); 4$\times$({\bf 1,4,1}); 8$\times$({\bf 1,1,8}) \\
\hline
\end{tabular}
\vskip 0.6cm
{Table 1:  The K3 orientifolds of type A \cite{GJ}, their gauge groups
living on D9-branes and corresponding charged hypermultiplets. Our
notation for the Chan-Paton twist matrix is explained in the text.}
\vskip 0.9cm

%%%%%%%%%%%%%%%%%%%%%%%%%%%%%%%%%%%%%%%%%%%

The expressions (\ref{u07}) and 
(\ref{f777}) for the one-loop corrections
 to the gauge couplings, obtained at the special symmetric points in
moduli space,  can be in fact generalized
easily to any $K3\times T^2$ orientifold  of type-I string
theory. One needs only to replace 
\be
b_a \to  2\sum_r T_a(r) e^{-\pi t M^2_r}- 2 T_a(G)
\ee
in (\ref{f777}), 
with $M_r$ the masses  of the hypermultiplets in six dimensions, and
modify the Kaluza-Klein sum appropriately if there are non-vanishing
Wilson lines on the two-torus. The  potential
$t\to 0$ divergence of the integral, and hence also the subtraction
term in (\ref{sub}),  are not affected by these `soft' corrections.

%%%%%%%%%%%%%%%%%%%%%%%%%%%%%%%%%%%%%%%%%%%

Let us turn now to  the ultraviolet divergences of the
amplitudes (3.3-3.5). To investigate them  we must reexpress the
amplitudes  as integrals over the
 proper time $l$ in the closed-string channel. Performing the sequences
(\ref{u9},\ref{u10})  of modular
transformations, a Poisson resummation of the
Kaluza-Klein sum, 
and using the well-known modular properties of the Dedekind
and Jacobi functions (see appendix A)
one finds after some algebra

\be\eqalign{
{\cal A}_{99}^{(k)}(B) = -
 Bl\; {  W^{(2)}(l) \over 16 \pi^3 }\; {\sqrt G} 
 \sum_{\a,\b=0,1/2} \eta_{\a,\b} \frac{\vartheta[{\b \atop \a }]}{\eta^3}&\ 
 {\rm tr}\left( (Q\g_9^k\otimes \g_9^k +
\g_9^k\otimes Q \g_9^k)
 \frac{\vartheta[{\b \atop \a}]({ \e}) }{\vartheta[{1/2 \atop 1/2
     }]({ \e})}
\right) \cr\times & 
 (2\sin{\pi k\over N} )^2
\frac{
\vartheta[{{\b + k/N} \atop \a}] \vartheta[{{\b - k/N} \atop \a}]}
 {  \vartheta[{1/2+k/N \atop 1/2 }]
\vartheta[{1/2 -k/N\atop 1/2}]}\ ,  \cr}
\ee
\be\eqalign{
{\cal M}_{99}^{(k)}(B) = 
 Bl\; {  W^{(2)}(4l) \over  \pi^3 }\; {\sqrt G} 
 \sum_{\a,\b=0,1/2} \eta_{\a,\b} \frac{\vartheta[{\a \atop \b }]}{\eta^3}&\ 
 {\rm tr}\left( Q\g_9^{2k} 
 \frac{\vartheta[{\a \atop \b}]({ \e\over 2}) }{\vartheta[{1/2 \atop 1/2
     }]({ \e\over 2})}
\right) \cr\times & 
 (2\sin{\pi k\over N} )^2
\frac{
\vartheta[{{\a + 2k/N} \atop \b+k/N}] \vartheta[{{\a - 2k/N} \atop \b-k/N}]}
 {  \vartheta[{1/2+2k/N \atop 1/2+k/N }]
\vartheta[{1/2 -2k/N\atop 1/2-k/N}]}\ ,  \cr}
\ee
\be
\eqalign{
{\cal A}_{95}^{(k)}(B) = 
 Bl\; {  W^{(2)}(l) \over 8 \pi^3 }\; {\sqrt G} 
 \sum_{\a,\b=0,1/2} \eta_{\a,\b} \frac{\vartheta[{\b \atop \a }]}{\eta^3}\ 
 {\rm tr}& \left( (Q\g_9^k\otimes \g_5^k)
 \frac{\vartheta[{\b \atop \a}]({ \e}) }{\vartheta[{1/2 \atop 1/2}]({ \e})}
\right) \cr\times & 
\frac{
\vartheta[{{\b + k/N} \atop \a+1/2}] \vartheta[{{\b - k/N} \atop \a+1/2}]}
 {  \vartheta[{1/2+k/N \atop 0 }]
\vartheta[{1/2 -k/N\atop 0}]}\ .
  \cr}
\ee

\noindent Taking  the 
 limit $l \rightarrow
\infty$ yields the expressions 
\be\eqalign{
\ \ \ \ {1\over l}{\cal A}_{99}^{(k)}(B) \;& \simeq \;
 {  B\; {\sqrt G} \over 2 \pi^3 }\;   \sin^2{\pi k\over N} \ 
{\rm tr}\left(  (Q\g_9^k\otimes \g_9^k +
\g_9^k\otimes Q \g_9^k) \Bigl[ {1\over {\rm sin}\pi\e }  -   {\rm cot}\pi\e    \Bigr] 
\right)\ ,\cr
\ \ \ \ {1\over l}{\cal M}_{99}^{(k)}(B)&  \;  \simeq\; 
- {  8B\; {\sqrt G} \over  \pi^3 }\;   \sin^2{\pi k\over N} \
{\rm tr}\left(  Q\g_9^{2k}\; \Bigl[ {1\over {\rm sin}(\pi\e/2) }  - 
  {\rm cot}(\pi\e/2)    \Bigr] \right) \ , \cr
\ \ \ \ \ {1\over l}{\cal A}_{95}^{(k)}(B) & \; \simeq\;
{  B\; {\sqrt G} \over 4 \pi^3 }\; {\rm tr}(\g_5^k)\;
{\rm tr}\left(  Q\g_9^{k}\; \Bigl[{1\over {\rm sin}\pi\e }  -   {\rm cot}\pi\e    \Bigr] 
\right)\ , 
\cr} \label{uvv}
\ee
where we have separated inside the square brackets the contributions
coming from the exchange of NS-NS closed-string states (proportional to the inverse
sines) and those coming from RR states (which are proportional to the
cotangents).

  Out of the above three amplitudes only ${\cal A}_{99}$ contains `mixed
  terms' with  charge-operator insertions at both the left and the
  right cylinder boundaries. The ${\cal A}_{95}$ diagram has one of it
  boundaries stuck on D5-branes,  which are blind to the (99) gauge
  groups, while the M{\"o}bius diagram has only a single boundary anyhow. These
  latter two diagrams combine with the `pure terms' (all charge
  insertions on the same boundary) of ${\cal A}_{99}$ to give an
  infrared-finite expression in the closed-string channel. This is a
  consequence of tadpole cancellation, which garantees that a
  zero-momentum massless particle cannot disappear into the vacuum in
  the absence of the magnetic-field background. This can be checked
  explicitly using the Chan-Paton matrices of table 1. For instance for
 the $Z_3$ model, there are  no D5-branes and  ${\rm tr}\g_9 = {\rm
  tr}\g_9^2 = 8$, which is precisely the condition for the sum of
  ${\cal A}_{99}^{(2k)}$ and ${\cal M}_{99}^{(k)}$ to be finite in the
  $l\to\infty$ limit of the integration.

    The leading non-cancelled divergences arise at quartic level from
    `mixed terms' of the ${\cal A}_{99}$ amplitudes. They are due to
    the exchange of twist-field scalars transforming in tensor
    multiplets of ${\cal N}=1$ supersymmetry in six dimensions, and
    coupling to the Yang-Mills action as in (\ref{ym}). All models
    with the exception of  $Z_2$ contain such tensor multiplets,
    localized at the fixed points of the orbifold \cite{GJ}. Since the
    perturbative heterotic string has only a single tensor multiplet,
    only  the $Z_2$ model has a  perturbative heterotic dual, whose threshold
    calculation agrees  with the results on  the type-I side \cite{ABFPT}.
Note  that supersymmetry does not allow the coupling (\ref{ym}) for
    hypermultiplets, so that in the $Z_2$ model the twisted cylinder amplitude
    is  infrared finite \cite{BF}.

 Expanding out to quartic order the expressions (\ref{uvv}) and using
 (\ref{u6}) we find
\be
\Lambda_4 = - {12 \pi^4 {\sqrt G} \over N} \sum_{k=1, k \not=N/2}^{N-1} \sin^2\! \pi kv
({\rm tr} Q^2 \g^k)^2 \int dl \ . \label{f12}
\ee

The physical interpretation of the term (\ref{f12}) of the type $({\rm tr} F^2)^2$ is 
that twisted NS-NS fields $m_k$ appear in the tree-level (disk) gauge kinetic function of
the gauge  group and generate at one-loop (tree-level in the transverse, closed string
picture) a  tadpole. By using
the integral form (\ref{f120}) for the propagator of  a       
canonically-normalized scalar\footnote{The gauge fields couple actually to linear
combinations of twisted fields which in our conventions are canonically normalized}, one
can extract the couplings
$s_k$ (\ref{ym}) for the various models. The result is:
\be
s_k =  {2 \pi^2 \over G^{1/4} \sqrt{N \pi}}
 | \sin {\pi kv}| ( {\rm tr} Q^2 \g^k)  \ . \label{f13}
\ee 
Because of the Peccei-Quinn symmetries generated by the
RR axions, these couplings\footnote{Strictly speaking, the metric $g_{kl}$ of $m_k$
should appear also in (\ref{f13}). The string result (\ref{f13}) is
valid around $m_k=0$, it therefore contains only the first term
$g_{kl}(m)=\delta_{kl}+ \cdots$ in a Taylor expansion.} must be linear 
(at the perturbative level) in $m_k$.

As previously discussed, these couplings are zero for the $Z_2$ example
in Table 1 and for the corresponding sectors $k=N/2$ of the models $Z_4$
and $Z_6$. They are also zero for the $k=1$ twisted moduli couplings to
the $U(8)$ gauge factor in the $Z_6$ model, because ${\rm tr} Q_{U(8)}^2
\g_9=0$ in this case. 

In one particular example ($Z_3$) these couplings are proportional to
the 4d beta functions and therefore in this case the unification is preserved.
Notice however that the twisted moduli fields are exact
flat directions in the effective field theory, therefore by using the
expression of the gauge couplings (\ref{ym}) we find that the
unification scale is an arbitrary parameter in these models.  In all the
other examples, unification is lost because the couplings $s_k$ are not
proportional to the 4d beta functions.
We should also mention that some
of the gauge couplings become strong for critical values of $m_k$, as it was
first pointed out in \cite{S}.

%%%%%%%%%%%%%%%%%%%%%%%%%%%%%%%%%%%%%%%%%%%%%%%%%%%%%%%%%%%%%%%%%%%%%%%%%%%%%%%%%%%%%%%
\section{ ${\cal N}=1$ supersymmetry: the $Z_3$ and $Z_7$ models}

 Our goal in this section is to compute 
the one-loop corrections to the gauge coupling constants coming from the ${\cal N}=1$ sectors.
in $Z_N$ orbifolds with $N$ a prime integer. Due to the absence of order
two twist elements, these orientifolds have no 5-branes in the spectrum
and are therefore the simplest 4d models with  ${\cal N}=1$ supersymmetry.
  
The annulus amplitude in the $Z_N$ type I orientifolds with odd $N$ can be written
\be
{\cal A} = {\cal A}_{{\cal N}=4}- \frac{1}{2N} \sum_{k=1}^{N-1}  \int_0^\infty \, \frac{dt}t \,
 \, {\cal A}^{(k)} (q) \ , \label{d1} 
\ee
where ${\cal A}_{{\cal N}=4}$ is the contribution of the ${\cal N}=4$
supersymmetric open spectrum, which 
does not contribute to the threshold corrections to the
gauge couplings and  ${\cal A}^{(k)}$ is the contribution of the 
${\g}^k \equiv ({\g})^k$ sectors given by
\be
{\cal A}^{(k)} = {1 \over 8 \pi^4 t^2} \sum_{\a,\b=0,1/2}  \eta_{\a,\b} \
\frac{\vartheta[{\a \atop \b}]}{\eta^3} \ \prod_{i=1}^3 (-2\sin \pi k v_i)
\  \frac{\vartheta[{\a \atop {\b + kv_i}}]}{\vartheta[{1/2 \atop {1/2 +kv_i}}]} \,  \
( {\rm tr} {\g}^{k})^2 \ . \label{d2}
\ee
The M{\"o}bius amplitude can be similarly written as in (\ref{d1}) by substituting
${\cal A} \rightarrow {\cal M}$, with
\be
{\cal M}^{(k)} = -  {1 \over 8 \pi^4 t^2} \sum_{\a,\b=0,1/2}  \eta_{\a,\b} \
\frac{\vartheta[{\a \atop \b}]}{\eta^3} \ \prod_{i=1}^3 (-2\sin \pi k v_i)
\  \frac{\vartheta[{\a \atop {\b + kv_i}}]}{\vartheta[{1/2 \atop {1/2 +kv_i}}]} \
({\rm tr} {\g}^{2k}) \ . \label{d3}
\ee
Because of supersymmetry, the amplitudes (\ref{d2}), (\ref{d3})
vanish in the absence of the magnetic field.

In the presence of the background magnetic field $B$, by using the modification 
(\ref{u4}), (\ref{u5}), the two amplitudes become
\ba
{\cal A}^{(k)}(B) \!\!&\!\!=\!\!&\!\! {iB \over 8 \pi^3 t} {\rm tr} \!
\left( \! (Q \g^k \! \otimes \! \g^k \!+\! \g^k \! \otimes Q \g^k)
\!\!\!\! \sum_{\a,\b=0,1/2}  \!\! \eta_{\a,\b} \
\frac{\vartheta[{\a \atop \b }]({i \! \e t \over 2})}{\vartheta[{1/2 \atop
1/2 }]({i \! \e t \over 2})} \right) \!\! 
\prod_{i=1}^3 (\!-\!2\sin \pi k v_i)
\  \frac{\vartheta[{\a \atop {\b + kv_i}}]}{\vartheta[{1/2 \atop 
{1/2 + kv_i}}]}  
 \ , \nonumber \\
{\cal M}^{(k)}(B) \!\!&\!\!=\!\!&\!\! -\!{ iB \over 4 \pi^3 t} {\rm tr} \left( Q
\g^{2k} \!\! \sum_{\a,\b=0,1/2} \!\! \eta_{\a,\b} \
\frac{\vartheta[{\a \atop \b }]({i \e t \over 2})}{\vartheta[{1/2 \atop
1/2 }] ({i \e t \over 2})} \right) 
\prod_{i=1}^3 (\!-\!2\sin \pi k v_i)
\  \frac{\vartheta[{\a \atop {\b + kv_i}}]}{\vartheta[{1/2 \atop {1/2 + kv_i}}]}  
 \ . \label{d4}
\ea
In computing the threshold corrections, we are interested in the quadratic $B^2$ terms
in a weak-field expansion.  By using the identity ${\rm tr} Q \g^k=0$
for a nonabelian gauge factor and after a straightforward algebra, we find 
(${\cal A} \equiv {\cal A}_0 + B^2 {\cal A}_2 + \cdots$, similarly for
${\cal M}$)
\ba
{\cal A}_2^{(k)} \!\!&\!\!=\!\!&\!\! {-B^2 \over 32 \pi^4} {\rm tr} (Q^2
\g^k \otimes \g^k) 
\int {dt \over t} \sum_{\a,\b=0,1/2}  \eta_{\a,\b} \
\frac{\vartheta^{''}[{\a \atop \b }]}{\eta^3 } 
\prod_{i=1}^3 (-2\sin \pi k v_i)
 \frac{\vartheta[{\a \atop {\b + kv_i}}]}{\vartheta[{1/2 \atop {1/2 + kv_i}}]}  
 \ , \nonumber \\
{\cal M}_2^{(k)} \!\!&\!\!=\!\!&\!\! {B^2 \over 16 \pi^4} {\rm tr} (Q^2 \g^{2k}) 
\int {dt \over t} \sum_{\a,\b=0,1/2}  \eta_{\a,\b} \
\frac{\vartheta^{''}[{\a \atop \b }]}{\eta^3} \ 
\prod_{i=1}^3 (-2\sin \pi k v_i)
\  \frac{\vartheta[{\a \atop {\b + kv_i}}]}{\vartheta[{1/2 \atop {1/2 + kv_i}}]}  
 \ . \label{d6}
\ea 
It is useful in the following to use the modular identity formula, valid
for $v_1+v_2+v_3=0$ (see Appendix A)
\be
 \sum_{\a,\b=0,1/2}  \eta_{\a,\b} \
\frac{\vartheta^{''}[{\a \atop \b }]}{\eta^3} \ 
\prod_{i=1}^3  \  
\frac{\vartheta[{\a \atop {\b + kv_i}}]}{\vartheta[{1/2 \atop {1/2 + kv_i}}]}  
= -2 \pi \sum_{i=1}^3 \frac{\vartheta^{'} [{1/2 \atop {1/2 - kv_i}}]}
{\vartheta [{1/2 \atop {1/2 - kv_i}}]} \ . \label{d8}
\ee
By using (\ref{d8}) into (\ref{u1}), (\ref{u6}) and (\ref{d6}), we
arrive at the final result for the one-loop threshold corrections
\be
{1 \over g^2}={1 \over g_0^2} - {1 \over 8 \pi N} \sum_{k=1}^{N-1} \int {dt \over t}
\prod_{i=1}^3 (-2\sin \pi k v_i) [( {\rm tr} Q^2 \g^k)( {\rm tr}
\g^k)-2( {\rm tr} Q^2 \g^{2k})]
 \sum_{j=1}^3 \frac{\vartheta^{'} [{1/2 \atop {1/2 - kv_j}}]}
{\vartheta [{1/2 \atop {1/2 - kv_j}}]} \ . \label{d9}
\ee
A first check of the formula (\ref{d9}) is by taking the infrared limit $t \rightarrow
\infty $ , in which case the one-loop expression (\ref{d9}) must reproduce the
one-loop running of the effective field theory, controled by the renormalization
group (RG) coefficients. In this limit we find
\be
\lim_{q \rightarrow 0}  
\frac{\vartheta^{'} [{1/2 \atop {1/2 - kv_j}}]}
{\vartheta [{1/2 \atop {1/2 - kv_j}}]} =  {\pi \cos (\pi k v_i) \over \sin (\pi k v_i)}
\ . \label{d10}
\ee

Let's now check the result (\ref{d9}) in the $Z_3$ example case.
The $Z_3$ four-dimensional ${\cal N}=1$ type I orientifold \cite{ABPSS} is defined by the twist
vector ${\bf v} = (1/3,1/3, -2/3)$. The tadpole consistency conditions ask for $32$ D9
branes in the spectrum and fix the Chan-Paton
matrix $\g = diag \ (  e^{2 i \pi /3} I_{12}, I_4 )$, where $I_N$ is the
$N \times N$ identity matrix. The matrix $\g$
then determines the gauge group to be $SU(12) \times SO(8) \times U(1)_X$ (the 
$U(1)_X$ factor is anomalous \cite{ABPSS}, \cite{K}) and the charged matter fields
are in the representations $3 {\bf (12,8)}_{1}+ 3 {\bf ({\overline
{66}},1)}_{-2}$, where the subscripts denote the $U(1)_X$ charges. This model contains
only ${\cal N}=4$ ($\theta^0$) and ${\cal N}=1$ ($\theta^1$,
$\theta^2$ )sectors.

In this example we compute
\be
( \prod_{i=1}^3 \sin \pi k v_i )  \sum_{j=1}^3 
{\cos (\pi k v_j) \over \sin (\pi k v_j)}= - {9 \over 8} \ , \label{d11}
\ee
for $k=1,2$ and, for $Z_3$ we use ${\rm tr} \g = -4$. We choose the generator $Q$  such
that, in a $U(16)$ complex basis it reads 
\be
Q_{SU(12)}={1 \over 2} \ diag \ (1,-1,0^{14}) \ \ , \ \ Q_{SO(8)}= Q_{U(4)} = {1 \over 2} 
\ diag \ (0^{12},1,-1,0,0) \ , \label{d12}
\ee
and therefore ${\rm tr} Q^2 \g^k = -1/2$ for $SU(12)$ and
${\rm tr} Q^2 \g^k = +1$ for the $SO(8)$ gauge group factors.
Finally, by cutting-off the integral in (\ref{d9}) by introducing the infrared (IR) 
regulator $t \le 1/ {\mu}^2$, we find the IR behaviour
\be
{4 \pi^2 \over g_{SU(12)}^2} =  {4 \pi^2 \over g_{SU(12),0}^2}+{9 \over 2} 
\ln { \mu \over M_I}
\ , \ {4 \pi^2 \over g_{SO(8)}^2} =  {4 \pi^2 \over g_{SO(8),0}^2} - 
9 \ln { \mu \over M_I}
\ , \label{d13}
\ee
which is indeed in agreement with the field theoretical RG coefficients
$ b_1 (SU(12))=-9 \ , \ b_2 (SO(8))= 18$, where $b_a = -3 T_a (G)+\sum_r T_a(r)$.
Notice that the corrections (\ref{d9}) are independent of the compactification radii,
in analogy with the threshold corrections of ${\cal N}=1$ sectors of heterotic models 
\cite{DKL}.

The second check of consistency of (\ref{d9}) is to go into the transverse channel
and check that the UV divergences cancel. By using the modular transformation
(\ref{f8}) in (\ref{d4}), we find the amplitudes in the transverse channel
\ba
{1 \over l} {\cal A}^{(k)}(B) \!\!&\!\!=\!\!&\!\! {iB \over 8 \pi^3} \! 
{\rm tr} \! \left( \! (Q \g^k \otimes \g^k \!+\! \g^k \otimes Q \g^k) 
\!\!\! \sum_{\a,\b=0,1/2}  \!\! \eta_{\a,\b} \
\frac{\vartheta[{\a \atop \b }] (\e)}{\vartheta[{1/2 \atop 1/2 }](\e)}
\right) \!\! 
\prod_{i=1}^3 (\!-\!2\sin \pi k v_i)
 \frac{\vartheta[{{\a+kv_i} \atop \b }]}{\vartheta[{1/2 +kv_i \atop 1/2}]}  
 \ , \nonumber \\
{1 \over l}{\cal M}^{(k)}(B) \!\!&\!\!=\!\!&\!\! \!- { iB \over \pi^3}
\! {\rm tr} \left( Q \g^{2k} 
\!\! \sum_{\a,\b=0,1/2} \!\! \eta_{\a,\b} \
\frac{\vartheta[{\a \atop \b }]({\e \over 2})}{\vartheta[{1/2 \atop 1/2
}]({\e \over 2})} \right) 
\prod_{i=1}^3 (\!-\!2\sin \pi k v_i)
\  \frac{\vartheta[{\a +2kv_i \atop {\b + kv_i}}]}{\vartheta[{1/2 +2kv_i 
\atop {1/2 + kv_i}}]} \ . \label{d15}
\ea
The UV behaviour of the above amplitudes can be easily worked out by taking the $l
\rightarrow \infty$ limit in (\ref{d15}). We find, by defining 
$\eta_k = sign (\prod_{i=1}^3 (\sin \pi k v_i))$, 
\ba
{1 \over l} {\cal A}^{(k)}(B) \!\!&\!\!=\!\!&\!\! - {iB \over 8 \pi^3} 
\prod_{i=1}^3 (\!-\!2\sin \pi k v_i) {\rm tr} \left( (Q \g^k \otimes \g^k + \g^k \otimes Q \g^k) 
 \ [1 + i \eta_k ({1 \over \sin{\pi \e}}- \cot{\pi \e})] \right)
 \ , \nonumber \\
{1 \over l} {\cal M}^{(k)}(B) \!\!&\!\!=\!\!&\!\!\!{iB \over \pi^3} 
\prod_{i=1}^3 (\!-\!2\sin \pi k v_i) {\rm tr} \left( 
Q \g^{2k} \ [1 - i \eta_k ({1 \over \sin{\pi \e}}- \cot{\pi \e})] \right) \ , \label{d16}
\ea
where, as in the previous section, we have explicitly displayed the contributions
coming from the exchange of NS-NS and the RR  states. By performing a
small $B$ expansion in (\ref{d16}), it can be checked
that the UV divergence in the $B^2$ term 
cancels, as expected, if the tadpole consistency condition is imposed.
More explicitly, the quadratic UV divergence cancels provided
\be
{\rm tr} \g^{2k} = {4 \over 8  \prod_{i=1}^3 \cos \pi k v_i } \ , \label{d016}
\ee
equivalent to the usual tadpole condition for $Z_N$ odd
orbifolds ${\rm tr} \g^{2k} \!\!=\!\! {32 \prod_{i=1}^3 \cos \pi k v_i }$ if we use
the equality $64 ( \prod_{i=1}^3 \cos \pi k v_i )^2=1$. This last
equality is a consequence of the fact that the number of fixed points
for a $Z_N$ odd orbifold $N_k = 64 ( \prod_{i=1}^3 \sin \pi k v_i )^2$
is independent of $k$, in particular $N_k=N_{2k}$. 

 An interesting phenomenon appears by expanding the above expressions (\ref{d16}) 
at order $B^4$. Indeed, by a straightforward computation we then find
an UV divergence, equal to
\be
\Lambda_4 =  - {24 \pi^4 \over N } \sum_{k=1}^{N-1} 
( {\rm tr} Q^2 \g^k)^2 \prod_{i=1}^3 |\sin \pi k v_i|
\ \int dl \ , \label{d17}
\ee
where the terms $({\rm tr} Q^4 \g^k)$ cancel exactly between the annulus and
the M{\"o}bius.
The interpretation of this term of the type $(tr F^2)^2$ is that twisted NS-NS 
fields $m_k$ (the blowing-up modes
of the orbifold) appear at tree-level in the gauge kinetic function of the gauge group 
and generate at one-loop (tree-level in the transverse, closed string picture) a
tadpole. Therefore, the tree-level gauge couplings (\ref{ym}) become\footnote{Note
that going from eq. (\ref{d17}) to (\ref{d18}), there is a global sign ambiguity which
cannot be fixed by our computation but does not affect our conclusions. This ambiguity
propagates also in eqs. (\ref{d21}) and (\ref{t060}).}
\ba
{4 \pi^2 \over g_{a,0}^2} &=& {1\over {\ell}} +
\sum_{k=1}^{[{N-1 \over 2}]}s_{ak} m_k \label{u12}\\ 
&=& {1 \over \ell} +   \sum_{k=1}^{[{N-1 \over 2}]}
{8 \pi^2 \over \sqrt{2 \pi N}}
( {\rm tr} Q^2 \g^k) |\prod_{i=1}^3 \sin \pi k v_i |^{1/2} m_k \ , \label{d18}
\ea 
${\ell}=e^{\phi_4} v^{-1/2}$, with $\phi_4$ the 4d dilaton and $v$ the volume
of the 6d compact space in string units; its partner is the universal axion
$a^{\rm RR}$,  dual to the untwisted RR antisymmetric tensor. 
The relation (\ref{d18}) confirms the result conjectured \cite{IRU} on the basis
of spacetime supersymmetry and anomaly considerations. Notice that , even if the
one-loop threshold corrections in the IR (\ref{d13}) and the couplings
(\ref{d18})  are separately non-universal (gauge-group dependent), remarkably
enough the coefficients of the coupling of $m_k$ to the gauge fields are
proportional to the beta function coefficients \cite{IMR}, shifting in a
universal way the string scale for $Z_N$ odd orbifolds.
 
 The anomalous $U(1)_X$ gauge factor has peculiar properties. Fist of
all, notice that the result (\ref{d18}) applies to $U(1)_X$ as
well. Moreover, by
introducing a background magnetic field $B'$ for it, coupled to the
gauge group generator $Q_{U(1)_X} \equiv Q_X$, where for example $Q_X=(1^{12},0^4)$
in the case of the $Z_3$ orientifold,      
we find a quadratic divergence
\be
 {B'^2 \over 4N \pi^2} \sum_{k=1}^{N-1} \prod_{i=1}^3 | \sin \pi k v_i|
({\rm tr} Q_X \g^k)^2 \int \ dl \ , \label{d20}
\ee
which is physically interpreted as a mixing between the
$U(1)_X$ gauge field and the Ramond-Ramond axions (antisymmetric tensors
$C_{\mu \nu}^k$ in the string basis) in the tree-level lagrangian \cite{DGM}. In order
to identify the coupling, we use the (gauge-fixed) propagator
\be
\Delta^{\mu \nu, \rho \sigma} (k^2) \equiv <C^{\mu \nu} C^{\rho \sigma}> = 
(g^{\mu \rho} g^{\nu \sigma}- g^{\mu \sigma} g^{\nu \rho}){i \over k^2} \ ,\label{d200}
\ee
for the (RR) antisymmetric moduli. The resulting coupling at the orbifold point
$m_k=0$ is then
\be
- {1 \over 2 \sqrt{2N \pi^3}} \sum_{k=1}^{[{N-1 \over 2}]} 
\prod_{i=1}^3 | \sin \pi k v_i|^{1 \over 2}
(-i {\rm tr} Q_X \g^k) \ \e_{\mu \nu \rho \sigma} 
C_{\mu \nu}^k F^{\rho \sigma}_X \ . \label{d21}
\ee
The $U(1)_X$ gauge boson becomes massive breaking spontaneously the
symmetry, even for zero VEV's of the twisted fields $m_k$. However, the
corresponding global symmetry $U(1)_X$ remains unbroken, since the
Fayet-Iliopoulos terms vanish in the orbifold limit $m_k=0$ \cite{P}.  
This property might be used in order
to protect proton decay in low scale string models \cite{L}, \cite{ADD}, \cite{DDG}.

Notice that in the $Z_3$ case there is actually one linear (symmetric) combination of
twisted moduli (out of the $27$ blowing up modes) which couples to the
gauge fields, which is the same appearing in the mixing between the RR
axions and the anomalous $U(1)$.   

%%%%%%%%%%%%%%%%%%%%%%%%%%%%%%%%%%%%%%%%%%%%%%%%%%%%%%%%%%%%%%%%%%%%%%%%%%%%%%%%%%%%%%%%%
\section{ ${\cal N}=1$ supersymmetry:  $Z_6'$ and  general models}

The $Z_6'$ model is defined by the twist vector ${\bf v} = (1/6,-1/2,1/3)$. The tadpole
cancellation conditions ask for 32 D9 branes and one set of 32 D5 branes filling the
third compact coordinate. The D5 branes are considered here
to be all at the origin in the $(z_1,z_2)$ plane for simplicity, where
$(z_1,z_2,z_3)$ denote the three complex compact coordinates. The solution for the
Chan-Paton matrices reads
\cite{KS}
\be
\g_9 \ = \g_5 \ = \quad diag \ ( e^{i \pi /6} I_4, e^{5i \pi /6} I_4, i I_8 ) \ . 
\label{t1}
\ee
The gauge group of this model is $[U(4) \times U(4) \times U(8)]_9 \times
[U(4) \times U(4) \times U(8)]_5$ and the charged matter representations are
\ba
99\ {\rm or} \ 55 &:& 
({\bf 4},{\bf 4},{\bf 1})+({\bar {\bf 4}},{\bar {\bf 4}},{\bf 1})+({\bar 
{\bf 4}},{\bf 4},{\bf 1})+({\bf 6},{\bf 1},{\bf 1})+({\bf 1},{\bar {\bf 6}},{\bf 1})+
\nonumber \\ 
\ \ \ \ \ \ \ \ \  
&&({\bf 1},{\bf 1},{\bf 28})+ ({\bf 1},{\bf 1},{\overline {\bf 28}})+ 
({\bf 1},{\bf 4},{\bf 8})+({\bar {\bf 4}},{\bf 1},{\bar {\bf 8}})+
({\bf 4},{\bf 1},{\bar {\bf 8}}) + ({\bf 1},{\bar {\bf 4}},{\bf 8}) \ , \nonumber \\
59 &:& 
({\bf 1},{\bf 4},{\bf 1};{\bf 1},{\bf 4},{\bf 1})+({\bf 4},{\bf 1},{\bf 1};{\bf 1},
{\bf 1},{\bf 8})+({\bf 1},{\bf 1},{\bf 8};{\bf 4},{\bf 1},{\bf 1}) + \nonumber \\
&&({\bar {\bf 4}},{\bf 1},{\bf 1};{\bar {\bf 4}},{\bf 1},{\bf 1})+({\bf 1},{\bar{\bf 4}},
{\bf 1};{\bf 1},{\bf 1},{\bar {\bf 8}})+ 
({\bf 1},{\bf 1},{\bar {\bf 8}};{\bf 1},{\bar {\bf 4}},{\bf 1}) \ . \label{t2} 
\ea
This model contains the ${\cal N}=4$ sector $\theta^0$, ${\cal N}=1$ sectors coming
from $\theta,
\theta^5$ and ${\cal N}=2$ sectors coming from $\theta^2, \theta^3$ and $\theta^4$. 

Our goal is the computation of the one-loop corrections to the gauge couplings 
coming from the ${\cal N}=1$ and ${\cal N}=2$ sectors. Then we will comment on the
straightforward  generalization of these formulas for a generic four-dimensional
${\cal N}=1$ type I orientifold.  For this purpose, we choose the D9 brane gauge group
and therefore we introduce a  background  magnetic field coupled to the D9 branes. In
this case the relevant amplitudes to consider are ${\cal A}_{99}, {\cal A}_{95}$  and
${\cal M}_{9}$. The other choice of turning on a 
magnetic field in the D5 branes sector can be obtained easily from the
previous one by T-duality.
 
The one-loop open string amplitudes of this model with and without the background
magnetic field are displayed in the Appendix B. 
By collecting the results in Appendix B and performing a weak-coupling expansion
similar to the one in section 3, we find
\ba
&& 2N {\cal B}_a (t) = - \sum_{k=1,5} \prod_{i=1}^3 (-2 \sin {\pi k v_i})
[ ({\rm tr} Q_a^2 \g_9^k)({\rm tr} \g_9^k)-2 ({\rm tr} Q_a^2 \g_9^{2k})] \sum_{j=1}^3  
{1 \over \pi } \frac{\vartheta^{'} [{1/2 \atop {1/2 - kv_j}}]}
{\vartheta [{1/2 \atop {1/2 - kv_j}}]}  \nonumber \\
&& - 2 \sum_{k=2,4} 4 | \sin{\pi kv_1} \sin{\pi kv_3} |
[ ({\rm tr} Q_a^2 \g_9^k)({\rm tr} \g_9^k)-2 ({\rm tr} Q_a^2 \g_9^{2k})]
\ \Gamma_2^{(2)} \nonumber \\
&& + 2 \{ 32 ({\rm tr} Q_a^2 ) -4 \sin{3 \pi v_1} \sin{3 \pi v_2} 
[( {\rm tr} Q_a^2 \g_9^3) ({\rm tr} \g_9^3)-2 ({\rm tr} Q_a^2 \g_9^{6})] \} \ \Gamma_3^{(2)} 
\nonumber \\
&& + \sum_{k=1,2,4,5} ({\rm tr} Q_a^2 \g_9^k) ({\rm tr} \g_5^k) {2 \sin{\pi k v_3} \over \pi }
 \left ( \frac{\vartheta^{'} [{1/2 \atop {1/2 - kv_3}}]}
{\vartheta [{1/2 \atop {1/2 - kv_3}}]} + \sum_{i=1}^2 \frac{\vartheta^{'} 
[{0 \atop {1/2 - kv_i}}]}{\vartheta [{0 \atop {1/2 - kv_i}}]} \right )  \ . \label{t3}
\ea
In putting the final result in the form (\ref{t3}), we used also the modular identities
(\ref{a3}) in Appendix A. The contribution of the ${\cal N}=2$ sectors in (\ref{t3}) is
easily identified with the terms containing the Kaluza-Klein momenta
sums $\Gamma_i^{(2)}$ along the compact direction $z_i$.
Notice that the third line in (\ref{t3}) coming from the $k=3$ sector can be
identified with the threshold corrections in the $Z_2$ orbifold model \cite{BS}
discussed in \cite{BF}.

The generalization of the above-result to a generic ${\cal N}=1$ four-dimensional type I
orientifold is straightforward. A general such model contains in the spectrum $D5_i$
branes filling the 4d spacetime and a compact complex dimension $z_i$. Different
$\theta^k$ sectors have ${\cal N}=1$ and ${\cal N}=2$ supersymmetry, each one having a contribution
as in (\ref{t3}). The last line in (\ref{t3}) is replaced by the corresponding sum 
over $5_i$ brane contributions. 

Taking the infrared limit in the above expression (\ref{t3}) allows to give a general
formula for the beta function coefficient contributions of the various $\theta^k$ sectors.
By using the infrared limits (\ref{d10}) and (\ref{a4}) we find
\ba
&b_a& = {4 \over N} \sum_{k \not=N/2} [({\rm tr} Q_a^2 \g_9^k)({\rm tr}
\g_9^k)- 2({\rm tr} Q_a^2 \g_9^{2k})]
( \prod_{i=1}^3 \sin{\pi kv_i}) \sum_{j=1}^3 {\cos{\pi kv_j} \over \sin{\pi kv_j}}
\nonumber \\
&+& {1 \over N} \sum_{i, k \not= N/2} 
({\rm tr} Q_a^2 \g_9^k)({\rm tr} \g_{5_i}^k) \cos{\pi kv_i}
+{24 \over N} {\rm tr} Q_a^2 \ , \label{t4}
\ea
where we considered here an $Z_N$ orientifold, for simplicity and the last contribution 
in the right-hand side of (\ref{t4}) comes from ${\cal A}_{95}^{(0)}$
and ${\cal M}_{99}^{(N/2)}$. The second line in (\ref{t4}) exist only
for even $N$ orientifolds.
In the particular case
of the $Z_6'$ orientifold, by putting the generator $Q_a$ into different gauge algebra
subgroups, it can be indeed checked that (\ref{t4}) reproduces the beta function
coefficients of the effective field theory $b_1 (SU(4))=9, b_1 (SU(8))=-6$.
It is also straightforward to study the general
UV structure by taking the $l \rightarrow \infty $ limit 
of the various terms (\ref{a21})-(\ref{a26}) in the transverse channel.
The UV finiteness of (\ref{t3}) is then obtained by using the tadpole 
conditions ${\rm tr} \g_9^k= {\rm tr} \g_5^k=0$ for
$k=1,3,5$, ${\rm tr} \g_9^2= {\rm tr} \g_5^2= -8$, ${\rm tr} \g_9^4=
{\rm tr} \g_5^4= 8$.   
By applying the same method as in the $Z_3$ orientifold case studied in the previous 
section, by expanding the transverse amplitudes to the order $B^4$ we find an UV
divergence coming from the ${\cal N}=1$ sectors $k=1,5$  and ${\cal N}=2$ sectors
$k=2,4$ 
\be
\Lambda_4 = - {3 \pi^4 \over N } \left[ \sum_{k=1,5} 
( {\rm tr} Q_a^2 \g^k)^2 \prod_{i=1}^3 
|2\sin \pi k v_i| + \sum_{k=2,4} v_2 ( {\rm tr} Q_a^2 \g^k)^2 \prod_{i=1,3} 
|2\sin \pi k v_i| \right] \int dl \ , \label{t040}
\ee
where $v_2=\sqrt{G_2}/\alpha'$ is the volume in string units of the second
compact torus. The result (\ref{t040}) is interpreted as a tree-level
modification of the gauge kinetic functions (\ref{u12}) with
\ba
&&\sum_k s_{ak} m_k= {\sum_{k=1}^{[{N-1 \over 2}]}}^\prime {2 \pi^2 \over \sqrt{\pi N}}
({\rm tr} Q_a^2 \g^k) |{\prod^3_{i=1}} 2 \sin \pi k v_i |^{1/2} m_k \ , kv_i
\not= {\rm integer}
\nonumber \\
&&\sum_k s_{ak} m_k = {\sum_{k=1}^{[{N-1 \over 2}]}}^\prime {2 \pi^2 \over
\sqrt{\pi N v_{i_0}}}
({\rm tr} Q_a^2 \g^k) |{\prod_{i \not=i_0}} 2 \sin \pi k v_i |^{1/2}
m_k \ ,  kv_{i_0} = {\rm integer}
\ , \label{t060}
\ea
and the prime in the sum excludes the sectors $k$ with $2kv_i = {\rm integer}$ for all 
$i=1,2,3$, while $v_{i_0}$ is the volume of the ${\cal N}=2$ complex
torus $T_{i_0}$ in string units. 
The sectors $k$ excluded from the sum are associated to D5 branes
and the corresponding twisted RR moduli are 4-forms in six dimensions; they 
belong to (neutral) hypermultiplets, which cannot 
couple to the kinetic terms of non-abelian gauge fields, by virtue of ${\cal N}=1$ 
supersymmetry in 6d. The remaining sectors fall in two categories: (i) ${\cal N}=1$
sectors corresponding to nontrivial twists for all three planes with associated twisted 
moduli described by linear multiplets in 4d; (ii) ${\cal N}=2$ sectors with no associated
D5 branes and their corresponding (real) moduli belonging to tensor muliplets in six
dimensions containing also their RR 2-form counterpart. The moduli of both categories
(i) and (ii) can generally couple to gauge fields according to (\ref{u12}).

Let us illustrate this general result to the case of $Z_6'$ orientifold. In this example, 
one has $N=6$ in (\ref{t060}) and the prime in the sum excludes the $Z_2$ sector $k=3$ 
associated to 32 D5 branes. 

We now describe the threshold corrections coming from  ${\cal N}=2$ sectors in (\ref{t3}), 
using the results of section 3.
These corrections depend on the geometric moduli $T_i,U_i$  defined as in eq. (\ref{f14}) 
for the three complex planes \cite{AFIV}
\be
S=a^{RR}+i {\sqrt{G_1G_2G_3} M_I^6 \over \lambda_I} \ , \ 
U_i = {G_i^{12}+i {\sqrt G_i} \over G_i^{22}} \ , \ T_i = b_i^{RR} + i {{\sqrt G_i} M_I^2 
\over \lambda_I}
\  \label{t06} 
\ee 
where $G_i$ is the metric on the torus $T_i$, related to the corresponding volume
$v_i=\sqrt{G_i}M_I^2$ (see eq. (\ref{t040})). Then the threshold corrections
(\ref{u07}) in the direct (open string) channel are equal to
\be
\Lambda_{2,a} = {1 \over 12} \sum_i b_{ai}^{({\cal N}=2)} \int {dt \over
t} \sum_{(m_i^1,m_i^2)} \left[
4 \ e^{-{t \over {\sqrt G_i}{\rm Im}U_i} |m_i^1+U_i m_i^2|^2} - \ e^{-{t \over {\sqrt
G_i}{\rm Im}U_i} |m_i^1+U_im_i^2|^2} \right] \ , \label{t5}
\ee
where $b_{ai}^{({\cal N}=2)}$
is the effective theory beta function coefficient of the corresponding ${\cal N}=2$
sector.\footnote{Notice that our definition of $b_{ai}^{({\cal N}=2)}$ differs
from the definition of ref. \cite{DKL} in the sense that ours represents the 
contribution of the ith ${\cal N}=2$ sector to the total beta function and therefore
equals $b_{ai}^{({\cal N}=2)}/ind$ in their notation.} 
By explicitly computing (\ref{t5}), we find the result 
\ba
&&\Lambda_{2,a} = -{1 \over 4} \sum_i b_{ai}^{({\cal N}=2)} 
 \ln ( {\sqrt G_i} {\rm Im}U_i \mu^2) + {\rm Im} f_a^{(1)}= \nonumber \\
&& -{1 \over 4} \sum_i b_{ai}^{({\cal N}=2)} 
 \ln \left[ \left({{\rm Im}S \ {\rm Im}T_i \over {\rm Im} T_j \ {\rm Im} 
T_k}\right)^{1/2} \ {\rm Im}U_i 
{\mu^2 \over M_I^2} \right] + {\rm Im} f_a^{(1)}
\ , \label{t7}
\ea 
with $j \not= k \not= i$ and where 
\be
f_a^{(1)}(U)= - i  \sum_i b_{ai}^{({\cal N}=2)} \ln \eta(U_i) \ . \label{t8}
\ee
The corrections (\ref{t7}) are similar with the 
heterotic ones in the ${\rm Im T}_i \rightarrow \infty$ limit, taking
into account that on the heterotic side the complex
structure moduli have the same definition (\ref{t06}), while
\be
S = a + i {\sqrt{G_1G_2G_3} M_H^6 \over \lambda_H^2} \ , \ T_i=b_i + i
{\sqrt G_i} M_H^2 \ . \label{t9}
\ee
Unlike the type I case, only the universal axion $a$ is described by a linear multiplet,
the others fitting naturally into chiral multiplets. 

Let's consider now the corrections given by an ${\cal N}=2$ sector,
depending on the complex torus of radii $R_{1,2}$.
Notice that in the limit $R_1,R_2 \rightarrow \infty$
with $R_1/R_2 = {\rm Im}U$ fixed, $\Lambda_2 \sim \ln (R_1R_2 \mu^2)$, whereas in the limit
$R_1 \rightarrow \infty$, $R_2$ fixed, the corrections are linearly divergent
$\Lambda_2 \sim R_1/R_2$. These power-law corrections can be used for 
the phenomenological purpose of lowering the unification scale \cite{DDG} in models 
with a low value of the string scale $M_I$ \cite{L,ADD}.
Alternatively, if $R_1/R_2 \simeq 1$ the logarithmic running  
$\Lambda_2 \sim \ln (R_1R_2 \mu^2)$ can also be used in order to achieve unification at 
a high Kaluza-Klein scale, even if the fundamental string scale has much lower 
values \cite{B}.
%%%%%%%%%%%%%%%%%%%%%%%%%%%%%%%%%%%%%%%%%%%%%%%%%%%%%%%%%%%%%%%%%%%%%%%%%%%%%%%%%%%

\section{Threshold corrections in models with spontaneous ${\cal N}=4 \rightarrow
{\cal N}=2$  supersymmetry breaking}

Type I string models with a continous parameter which interpolates between models with
different numbers of supersymmetries are an interesting framework for study threshold
corrections and their dependence on the interpolating parameter.
In particular we show that in the limit where the maximal supersymmetry is restored,
there are no linear corrections in the corresponding radius. The dependence is logarithmic
only, which could be an advantage, in the sense that the gauge hierarchy
problem corresponding to these dimensions is improved \cite{AB}. 

{\large ${\cal N}=4 \rightarrow {\cal N}=2$ Scherk-Schwarz breaking}

This model can be described in the closed string sector , as a freely-acting orbifold 
$IIB/(-1)^m {\cal I}$, where
$(-1)^m$ denotes the order-two shift $X^5 \rightarrow X^5 + \pi R_1$ and ${\cal I}$
denotes the inversion of the four internal coordinates ${\cal I} X^{6 \cdots9}
= - X^{6 \cdots9}$ \cite{KK} (a $R_1 \rightarrow 2R_1$ operation is required in order to
go in the Scherk-Schwarz basis in which the amplitudes below are written) . The 
resulting type I model \cite{ADDS} has ${\cal N}=2$ supersymmetry in 4d and a
gauge group $SO(N_1) \times SO(N_2)$ (with $N_1+N_2=32$) originating from D9 branes 
and can be described as
a Scherk-Schwarz deformation (a shift of the Kaluza-Klein modes $m_1$ with $1/2$ unit
along $R_1$) of the ${\cal N}=4$ supersymmetric type I model with a Wilson line
that breaks $SO(32)$ down to $SO(N_1) \times SO(N_2)$.
${\cal N}=4$ supersymmetry is recovered in the $R_1 \rightarrow \infty$ limit.
The open string massless spectrum contains, besides the adjoint ${\cal N}=2$ vector multiplets 
${\bf (N_1(N_1-1)/2,1)} + {\bf (1,N_2(N_2-1)/2)}$, one hypermultiplet in the 
representation ${\bf (N_1,N_2)}$.

The open string amplitudes for vanishing magnetic field are given by the expressions
\ba
&&{\cal A} = {({\rm tr} 1)^2 \over 8 \pi^4 t^2} \sum_{\a,\b=0,1/2}  \eta_{\a,\b} \
\frac{\vartheta^4 [{\a \atop \b}]}{\eta^{12}} \Gamma^{(4)}
(\Gamma_{m_1}+ \Gamma_{m_1+1/2}) \Gamma_{m_2} +
\nonumber \\
&&{({\rm tr} \g)^2 \over 8 \pi^4 t^2} \sum_{\a,\b=0,1/2}  \eta_{\a,\b} \
\frac{\vartheta^2 [{\a \atop \b}]}{\eta^{6}} 
\frac{4 \vartheta[{\a \atop \b +1/2}] \vartheta[{\a \atop \b -1/2}] }{\theta_2^2} 
(\Gamma_{m_1}-\Gamma_{m_1+1/2}) \Gamma_{m_2} \ , \nonumber \\
&&{\cal M} = - {({\rm tr} 1) \over 8 \pi^4 t^2}   \sum_{\a,\b=0,1/2}  \eta_{\a,\b} \
\frac{\vartheta^4 [{\a \atop \b}]}{\eta^{12}} \Gamma^{(4)}
(\Gamma_{m_1}+\Gamma_{m_1+1/2}) \Gamma_{m_2} - \nonumber \\
&& {({\rm tr} 1) \over 8 \pi^4 t^2}  \sum_{\a,\b=0,1/2}  \eta_{\a,\b} \
\frac{\vartheta^2 [{\a \atop \b}]}{\eta^{6}} 
\frac{4 \vartheta[{\a \atop \b +1/2}] \vartheta[{\a \atop \b -1/2}] }{\theta_2^2} 
(\Gamma_{m_1}-\Gamma_{m_1+1/2}) \Gamma_{m_2} \ , \ \label{q1}
\ea
where the action of the twist on the Chan-Paton degrees of freedom is $\g = 
\ diag \ (I_{N_1}, -I_{N_2})$. In (\ref{q1}), $\Gamma_{m_1}(\Gamma_{m_2})$ denotes the Kaluza-Klein 
momentum sum along $X^5 (X^4)$ and $\Gamma^{(4)}$ denotes the momentum sum along $T^4$.

The introduction of the background magnetic field  
is completely analogous to the cases studied in the previous sections.  By using 
the relations ${\rm tr} 1 = N_1+N_2$, ${\rm tr} \g = N_1-N_2$, ${\rm tr} Q^2 \g = \pm 1$ 
(plus sign for $SO(N_1)$ and
minus sign for $SO(N_2)$ gauge factors) and by using the last modular identity in
(\ref{a2}), we find the threshold corrections
\be
{\cal B} (t) = -2 [ \pm (N_1-N_2)-2] (\Gamma_{m_1} - \Gamma_{m_1+1/2}) \ \Gamma_{m_2} \ , \label{q2}
\ee
which in the IR limit agree with the field theory beta-function coefficients
$b_1 (SO(N_1)) = -2N_1+2N_2+4$, $b_2 (SO(N_2)) = -2N_2+2N_1+4$.
Notice that, because of the remnant ${\cal N}=2$ supersymmetry, the string oscillators decoupled 
in the final expression, in analogy with the ${\cal N}=2$ sectors of orbifold models, as shown
in \cite{BF} and in section 3.

 These corrections, in analogy with eq. (\ref{t5}), can be written in a more
compact way in the transverse channel
\ba
&&\!\!\!\!\!\!\!\! \Lambda_2 \!\!=\!\! {-{\sqrt G} \over 2 \pi} \int_{\mu^2}^{\infty} dl  
\sum_{n_1,n_2} [1\!-\!(-1)^{n_1}] \left( \pm (N_1-N_2) e^{-{{\sqrt G} \over
{\rm Im}U} |n_2+Un_1|^2 l}\!-\! 8 
e^{-{4{\sqrt G} \over {\rm Im}U} |n_2+Un_1|^2 l} \right) \nonumber \\
&&= {1 \over 2} [\pm (N_1-N_2)-2] \ln [e^{-2 \gamma_E} \mu^2 \sqrt{G} {\rm Im}U 
|{\eta^3 (U) \over 2 \theta_2 (U)}|^2 ] - 2 \ln 2 \ , \label{q3} 
\ea
where $U$ is the complex field corresponding to the $R_1,R_2$ torus,
$\sqrt{G}=R_1R_2$ and $\gamma_E$ is the Euler number. In the $R_2
\rightarrow \infty$ limit, $R_1$ fixed  
we get $\Lambda_2 \sim R_2/R_1$ and in the $R_1,R_2 \rightarrow \infty$ limit we get
$\Lambda_2 \sim \ln (R_1 R_2)$, as expected. However, in the limit $R_1 \rightarrow
\infty$, $R_2$ fixed, corresponding to the limit of the restoration of the full ${\cal N}=4$
supersymmetry, we find just logarithmic corrections $\Lambda_2 \sim \ln R_1$,
as in the dual heterotic models studied in \cite{KKPR}. 
The corresponding
threshold corrections on the heterotic side agree with (\ref{q3}), as expected, in
the $T_H \rightarrow i \infty$ limit. 
Notice that in
(\ref{q3}) the UV divergence is automatically zero and does not ask for the untwisted
tadpole condition $N_1+N_2=32$, in analogy with the previous examples, because the untwisted
tadpole condition is related to the ${\cal N}=4$ sector which does not contribute to the
threshold corrections to the gauge couplings. The UV finiteness holds to all
orders in the magnetic field $B$, which means that no tree-level modification of the
gauge kinetic function appears here, which is to be expected since there are no fixed points
and therefore no twisted fields in a freely-acting orbifold.

{\large ${\cal N}=4 \rightarrow {\cal N}=2$ M-theory breaking}

This model is, in the closed string sector, the T-dual of the above one in the sense that
it exchanges the momentum modes and the winding modes along $R_1$. The deformation
shifts now the winding modes along $R_1$ with $1/2$ unit and ${\cal N}=4$ supersymmetry is
recovered in the $R_1 \rightarrow 0$ limit. The model, which can also be seen by duality
arguments as M-theory compactified on $(T^4 \times S^1)/(Z_2 \times Z_2') \times S^2$,
contains, by tadpole consistency conditions, $16$ D9 branes and $16$ D5 branes, with 
a gauge group $SO(16) \times SO(16)$. The massive spectrum of the model has ${\cal N}=2$ 
supersymmetry, but the massless one has ${\cal N}=4$ supersymmetry and consists of 
the vector multiplet in the adjoint representation of the gauge group. 
In order to distinguish the two gauge factors in the following, we write
the gauge group as  $SO(N) \times SO(D)$, where the tadpole
conditions ask for $N=D=16$.
   
The one-loop open string amplitudes for vanishing magnetic field read \cite{ADDS}
\ba
&&\!\!\!\!\!\!\!\!{\cal A} \!=\! {\cal A}_{{\cal N}=4} \!+\! {1 \over 2 \pi^4 t^2}
\!\! \sum_{\a,\b=0,1/2} \!\! \eta_{\a,\b} \!
\frac{\vartheta^2 [{\a \atop \b }]}{\eta^6}  
\frac{ \vartheta^2 [{\a \atop {\b + 1/2}}]}
{\vartheta[{0 \atop 1/2}]}  N D  \Gamma_{m_1+1/2} \Gamma_{m_2} \ , 
\nonumber \\
&&\!\!\!\!\!\!\!\!{\cal M} \!=\! {\cal M}_{{\cal N}=4} \!+\! {1 \over 4
\pi^4 t^2} \!\!\!\!
\sum_{\a,\b=0,1/2} \!\!\!\! \eta_{\a,\b} \! 
\frac{\vartheta^2 [{\a \atop \b }]}{\eta^6}  
\frac{4 \vartheta[{\a \atop {\b + 1/2}}] \vartheta[{\a \atop {\b - 1/2}}]}
{\vartheta[{1/2 \atop 0}]} (N + D) \Gamma_{2m_1+1} \Gamma_{m_2} \ , 
\label{q4}
\ea
where the ${\cal N}=4$ part is as usually irrelevant for our purposes.
We choose to turn on a magnetic field in the $SO(N)$ gauge group factor. By following the
same steps as in the previous sections we find the threshold corrections
\be
{\cal B} (t) = 2 [ -8 \Gamma_{2m_1+1}+D \Gamma_{m_1+1/2}] \Gamma_{m_2} \  
\label{q5}
\ee
and therefore there are no IR divergences. This is in agreement with the
vanishing of the beta function of the effective field theory.
The threshold corrections can be quantitatively computed and the result is
\ba
&&\!\!\!\!\!\!\!\! \Lambda_2 \!\!=\!\! {{\sqrt G} \over 2 \pi }
\int_0^{\infty} dl \sum_{n_1,n_2} 
 \left( \!(-1)^{n_1} D e^{-{{\sqrt G} \over {\rm Im}U}
|n_2+Un_1|^2 l}\!-\! 16 (-1)^{n_1} 
e^{-{4{\sqrt G} \over {\rm Im}U} |n_2+Un_1|^2 l} \right) \nonumber \\
&& \!\!\! \!=\!  
{D \over 2} \ln [4 |{\theta_2 (U) \over \eta (U)}|^2 ] +
4 \ln [4 |{\theta_2 (U/2) \over \eta (U/2)}|^2 ] \ . \label{q6} 
\ea 
In the limit of the restoration
of ${\cal N}=4$ supersymmetry $R_1 \rightarrow 0$, $R_2$ fixed, we find a result
exponentially suppressed $\Lambda_2 \sim e^{-R_2/R_1}$. This model has the property
that the tadpoles are locally cancelled in the compact direction $R_1$. This can be understood
by the fact that, in the transverse channel, the annulus and the M{\"o}bius amplitudes (\ref{q4})
have the same winding lattice sum $(-1)^n W_n$, while usually (in toroidal or orbifold 
compactifications without Wilson lines) the transverse annulus
contains the full lattice $W_n$, while the transverse M{\"o}bius contains only the even windings
lattice $W_{2n}$. Therefore not only the UV divergence
in (\ref{q6}) coming from the $n=0$ mode is cancelled once the tadpole 
condition $D=16$ is imposed, but also simultaneously the contribution of
the massive $n$ modes, which
describe the position in the transverse (to the branes) space.
The UV convergence holds at all orders, for the same physical
reason as in the model of the previous subsection.

%%%%%%%%%%%%%%%%%%%%%%%%%%%%%%%%%%%%%%%%%%%%%%%%%%%%%%%%%%%%%%%%%%%%%%%%%%%%%%%%%%%%%%%%%%%
\section{Effective field theory}

In the remaining two sections we discuss how our string-theory results
could fit (a) with the general expressions of supergravity and (b)
with the conjectured ${\cal N}=1$ heterotic/type I duality, although
there is no regime in which the 10d string coupling is small on both
sides. Thus, we should warn the reader that this discussion is only
tentative. Our main results will be (a) that the one-loop truncated supergravity
expressions can be fitted in the special cases of $Z_3$ and $Z_6'$
models and (b) that duality does not seem to work even in the $Z_3$
example where the perturbative spectra on the two sides agree. These
points deserve definitely further study.

The starting point is the general (all-loop order) expression
for the physical gauge couplings $g_a$ in locally supersymmetric field 
theories \cite{KL1}:
\ba
{4\pi^2\over g_a^2(\mu^2)}\!=\! {\rm Im} f_a\!+\!{b_a \over 4} \ln {M_P^2 \over \mu^2}
\!+\! {c_a \over 4} K \!+\! {T_a(G) \over 2} \ln g_a^{-2} (\mu^2)
\!-\! \sum_r {T_a(r) \over 2} \ln \det Z_{(r)}(\mu^2),
\label{e1}
\ea
expressed in terms of the Wilsonian (holomorphic) gauge couplings
$f_a$ and the wave-function normalization matrix  $Z_{(r)}$ for the charged 
matter fields.
Here, $M_P$ is the Planck mass, $K$ is the K{\"a}hler potential, $a$ denotes 
the gauge group factor and 
$r$ runs over the gauge group representations with Dynkin index $T_a(r)$.
The one-loop beta functions $b_a$ and the coefficients $c_a$ are given by
\be
b_a = \sum_r T_a(r) - 3 T_a(G) \ , \qquad   c_a = \sum_r T_a(r) -  T_a(G)  
\ . \label{e2}
\ee
Truncating the expression (\ref{e1}) to one-loop order requires the knowledge of
the holomorphic gauge couplings $f_a$ at one-loop, while the K{\"a}hler
potential and the wave-function normalization matrix need to be known only
at tree-level. For simplicity, in the following we concentrate on the gauge
couplings originating from D9 branes. The holomorphic gauge couplings 
and the K{\"a}hler potential can be generally expanded as
\ba
f_a &=& S + s_{ak} M_k +  f_a^{(1)} (U) \ , \nonumber \\
K &=& - \ln (S-{\bar S}) + {\hat K} (M_k,T,U) \ , \label{e3}
\ea
where $U$($T$) are the complex structure (K{\"a}hler class) 
moduli fields and $s_{ak}$ are numerical constants given by
(\ref{t060}). An important point is that in the Wilsonian gauge kinetic
function $f_a$ the fields $S,M_k$ form complex chiral multiplets and that
${\rm Im}S$, ${\rm Im}M_k$ are related to the linear multiplet fields
$\ell$, $m_k$ (\ref{d18}), (\ref{u12}) in the string basis through a
chiral-linear multiplet duality \cite{DFKZ}.
Notice that the twisted moduli $M_k$ can appear perturbatively only linearly 
in $f_a$ because of the Peccei-Quinn symmetries associated to the 
Ramond-Ramond axions. Similarly, the only possible dependence of $f_a$
on $T$ moduli is linear, but for D9 branes this dependence vanishes.
As we saw in the preceding sections, all linear dependence appears only
at the tree-level (disk diagram), while $f^{(1)}_a$ is the genus one correction (\ref{t8}).
Inserting eqs. (\ref{e2}) and (\ref{e3}) into (\ref{e1}), we find
\ba
&&{4\pi^2\over g_a^2(\mu^2)^{1-loop}}={\rm Im} S+ s_{ak} {\rm Im} M_k + {b_a \over 4}
(\ln {M_P^2 \over \mu^2}-\ln (S-{\bar S})) + \nonumber \\
&&{1 \over 4} \left[ 4 {\rm Im} f_a^{(1)}(U)+c_a {\hat K} -      
2 \sum_r T_a(r) \ln \det Z_{(r)} +2 T_a(G) \ln (1+ s_{ak} {{\rm Im} M_k 
\over  {\rm Im} S}) \right] \ . \label{e4}
\ea
On the other hand, as we discussed in the previous sections, a direct one-loop
type I string computation in orbifold compactifications gives the
(moduli-dependent) result
\ba
&&{4\pi^2\over g_a^2(\mu^2)^{1-loop}}= {1 \over \ell}+ s_{ak} m_k + \nonumber \\
&&{1 \over 4} \left[ 4 {\rm Im} f^{(1)}_a(U) + b_a^{({\cal N}=1)} \ln {M_I^2
\over \mu^2} - \sum_i b_{ai}^{({\cal N}=2)} \ln ({\sqrt G_i} {\rm Im} U_i \mu^2) \right] \ , 
\label{e5}
\ea 
where the total beta function coefficient is a sum of contributions of ${\cal N}=1$ and
${\cal N}=2$ sectors: $b_a=b_a^{({\cal N}=1)}+ \sum_i b_{ai}^{({\cal N}=2)}$.

Our goal here is to study the compatibility between the effective field theory
result (\ref{e4}) and the string theory result (\ref{e5}) in the models $Z_3$ and
$Z_6'$ discussed in the previous sections. We limit ourselves to the orbifold limit
$m_k \rightarrow 0$, where the string result is really valid. It is instructive
to review first the situation in the heterotic case. The corresponding effective
field theory and the heterotic string expressions can be obtained by
putting $s_{ak}=0$ and
${\rm Im} M_k=0$ in (\ref{e4}) and (\ref{e5}), replacing $M_I$ by the heterotic
string scale $M_H$ and allowing a dependence of the
analytic one-loop corrections $f_i^{(1)}$ on the K{\"a}hler class moduli $T$ \cite{KL2}.
This dependence turns out to be the sum ${\rm Im} f^{(1)}_a(T) + {\rm Im}
f^{(1)}_a(U)$. Indeed, unlike in type I strings, in heterotic strings the $T$ moduli
are not protected  by Peccei-Quinn symmetries. More explicitly, the string expression
becomes\footnote{Our notations are related to the function $\Delta_a$ of \cite{KL2}
through the relation
$\Delta_a = {\rm Im} f_a^{(1)}-\sum_i b_{ai}^{({\cal N}=2)} \ln ({\rm Im}T_i {\rm Im}U_i)$.}
 \ba
&&{4\pi^2\over g_a^2(\mu^2)^{1-loop}}= {k_a \over \ell} + {1 \over 4} Y (T,U) + \nonumber \\
&&{1 \over 4} \left[ 4 {\rm Im} f^{(1)}_a(T) +  {\rm Im} f^{(1)}_a(U) + b_a \ln {M_H^2
\over \mu^2} - \sum_i b_{ai}^{({\cal N}=2)} \ln ({\rm Im T}_i {\rm Im U}_i) \right] \ ,
\label{e6}
\ea 
where $k_a$ are the Kac-Moody levels and $Y$ is a universal correction coming
from the  ${\cal N}=2$ sectors (containing both the non-analytic and
analytic pieces computed in the literature \cite{DKL,DFKZ}). 
Notice first that the two terms inside the bracket 
multiplying $b_a$ in the first line of field theory expression (\ref{e4}) combine to form
$\ln (M_H^2/\mu^2)$, which reproduces the corresponding term in the string expression
(\ref{e6}). Furthermore, a case by case analysis shows that the remaining two terms
are equal to:
\be
c_a {\hat K}- 2 \sum_r T_a(r) \ln \det Z_{(r)} =
\sum_i (\delta_{\rm GS}^i- b_{ai}^{({\cal N}=2)}) \ln ({\rm Im T}_i {\rm Im U}_i)\, ,
\label{comp}
\ee
where $\delta_{\rm GS}^i$ are gauge group independent constants.
This apparent difference (by a universal term) between the two results can be
explained psrtly by the fact that the string result uses a linear multiplet for the dilaton $L$,
instead of a chiral one $S$ used in the field theory expression. 
Indeed, the duality transformation that changes basis from the string to the supergravity
framework, brings a universal term proportional to
the one-loop correction to the moduli K{\"a}hler potential \cite{DFKZ}. As
a result, the non-analytic part in $Y$ cancels
out while the coefficients $b_{ai}^{({\cal N}=2)}$ are shifted by the universal constants
$\delta_{\rm GS}^i$ as in eq. (\ref{comp}).

We present in detail the explicit example of the $Z_3$ orbifold of the $SO(32)$
heterotic string. The four-dimensional gauge group is $SU(12) \times SO(8) 
\times U(1)_X$ (the $U(1)_X$ factor is anomalous). 
The charged massless chiral multiplets are in the representations 
$3 {\bf(12,8)_{-1}}+ 3 {\bf ( \overline {66},1)_{+2}}$ from the untwisted sector
and $27 {\bf (1,1)_{-4}}+27{\bf (1,8_s)_{+2}}$ from the
twisted sector, where the subscripts denote the $U(1)_X$ charge. In addition there
are 9 neutral untwisted (K{\"a}hler class) moduli $T_{i \bar j}$. In the following, 
for simplicity we consider only the diagonal moduli $T_{i \bar i} \equiv T_i$. 
Since there are no ${\cal N}=2$ sectors, in the string expression (\ref{e6})
$\Omega=0, b_{ai}^{({\cal N}=2)}=0$ and there are no moduli
dependent corrections to the Wilsonian couplings $f^{(1)}=0$. On the other hand,
in order to evaluate the field-theory expression (\ref{e4}), we use the results
\ba
{\hat K} \!=\! - \sum_{i=1}^3 \ln (T_i-{\bar T}_i) \ , \ Z_{({\bf {12,8}})^i} \!=\! 
Z_{({\bf {{\overline 66},1}})^i}={1 \over {\rm Im} T_i} \ , \
Z_{({\bf {1,8_s}})} \!=\! {1 \over [({\rm Im}T_1) ({\rm Im}T_2) ({\rm Im}T_3)]^{2/3}}
 \ , \label{e7}
\ea
where the index $i$ in the untwisted matter labels the
three different representations. Notice that in (\ref{e4}), by using the
definitions (\ref{e2}) and (\ref{e7}), we can write
\be
c_a {\hat K} - 2 \sum_r T_a(r) \ln \det Z_{(r)} = -{1 \over 3} b_a \sum_i
\ln (T_i-{\bar T}_i) + {2 \over 3} \sum_{r \ \rm twisted} T_a (r) \sum_i \ln (T_i-{\bar T}_i)
\ , \label{e8}
\ee 
where the last sum in (\ref{e8})  is over the matter representations from the twisted
sector. Moreover, by using the relation $M_H^2 = M_P^2/({\rm Im}S)^2$, we conclude
that the string scale in (\ref{e4}) cancels between the first line and the 
untwisted sector contribution of the second line. Therefore (\ref{e4}) becomes,
for the two nonabelian gauge factors
\ba
{4 \pi^2 \over g_{SU(12)}^2} &=& {\rm Im} S - {b_{SU(12)} \over 12} \sum_i 
\ln (\sqrt{G_i} \mu^2)  \ , \nonumber \\
{4 \pi^2 \over g_{SO(8)}^2} &=& {\rm Im} S - {b_{SO(8)} \over 12} \sum_i 
\ln (\sqrt{G_i} \mu^2) + {1 \over 6} \sum_{r \ twisted } T_a(r) \sum_i \ln (T_i-
{\bar T}_i) \ , \label{e9}
\ea 
where the beta function coefficients are $b_{SU(12)}=-9$, $b_{SO(8)}=45$.
As a result, we can rewrite the field theory expressions (\ref{e9}) as the 
(moduli independent) string theory expressions up to a universal contribution
\ba
{4 \pi^2 \over g_{SU(12)}^2} &=& {\rm Im} S - {b_{SU(12)} \over 4}  
\ln ({\mu^2 \over M_H^2}) - {b_{SU(12)} \over 12} \ln (V M_H^6)  \ , \nonumber \\
{4 \pi^2 \over g_{SO(8)}^2} &=& {\rm Im} S - {b_{SO(8)} \over 4} 
\ln ({\mu^2 \over M_H^2}) - {b_{SU(12)} \over 12} \ln (V M_H^6) \ , \label{e10}
\ea
where $V= \sqrt{G_1 G_2 G_3}$ is the (dimensionful) volume of the compact space.
As discussed above, the universal term originates from the one-loop correction to
the moduli metric, which translates into a correction to the gauge couplings after
the change of basis from the string (linear multiplet) basis to the effective
supergravity (chiral multiplet) basis
\be
{1 \over \ell} = {\rm Im} S - \sum_i \delta_{GS}^i \ln (T_i-{\bar T}_i) \ , 
\label{e11}
\ee
where $\delta_{GS}^i$ are the so-called Green-Schwarz coefficients, which in
this case are equal to $ \delta_{GS}^1=\delta_{GS}^2=\delta_{GS}^3=b_{SU(12)}/12$.
Note that this phenomenon is absent in nonsingular (Calabi-Yau)
blown-up $Z_3$ compactifications
where no twisted states are present in the spectrum. In this
case, the last term in the second eq. in (\ref{e9}) is absent and the resulting
expression coincides with the large volume limit of the string result \cite{KL2}.
Notice that the heterotic string scale disappears and the compactification volume
plays the role of the unification scale.

We now turn to the type I case and start with the simplest $Z_3$ orientifold example
discussed in section 4. In this case, the string result is given by (\ref{e5})
with the couplings of twisted moduli proportional to the beta functions 
$s_{a}=c b_a$ ($a=SU(12),SO(8)$), $f_a^{(1)}=b_{ai}^{({\cal
N}=2)}=0$ due to the absence of ${\cal N}=2$ sectors, while $b_a^{({\cal N}=1)}$
are identical to the full beta functions $b_a$ given in (\ref{d13}). The result is
 \ba
{4 \pi^2 \over g_{SU(12)}^2} &=& {1 \over l} + c \ b_{SU(12)} \ m
-{b_{SU(12)} \over 4}  
\ln ({\mu^2 \over M_I^2})  \ , \nonumber \\
{4 \pi^2 \over g_{SO(8)}^2} &=& {1 \over l} + c \ b_{SO(8)} \ m
-{b_{SO(8)} \over 4}  \ln ({\mu^2 \over M_I^2})  \ . \label{e011}
\ea   
On the other hand, the field theory expression is given by (\ref{e4}), with the
functions ${\hat K}$ and $Z_{(r)}$ as in (\ref{e7}) but with the
representation ${\bf (1,8_s)}$ absent 
and where the complex moduli $T_i$ have the type I definitions (\ref{t06})). The
final result is
\ba
{4 \pi^2 \over g_{SU(12)}^2} &=& {\rm Im} S + c \ b_{SU(12)} \ {\rm Im} M
-{b_{SU(12)} \over 12} \sum_i 
\ln (\sqrt{G_i} \mu^2)  \ , \nonumber \\
{4 \pi^2 \over g_{SO(8)}^2} &=& {\rm Im} S + c \ b_{SO(8)} \ {\rm Im} M
-{b_{SO(8)} \over 12} \sum_i 
\ln (\sqrt{G_i} \mu^2)  \ . \label{e012}
\ea 
Therefore, the one-loop string theory and the
field theory results formally agree if the twisted moduli duality were
\be
m \ = \ {\rm Im} M - {1 \over 12 c} \ln (V M_I^6)
\ = \ {\rm Im}M - {1 \over 24 c} \ln {({\rm Im}S)^3 \over {\rm Im}T_1 
{\rm Im}T_2 {\rm Im}T_3 } 
\ . \label{e12}
\ee
This is a linear-chiral multiplet duality\footnote{For a recent paper
discussing duality in the case of several linear multiplets, see
\cite{BW}.} for the twisted moduli, analogous to (\ref{e11}), perturbatively
valid around $m_k=0$. The exact duality relation could depend on the
complete lagrangian for the twisted moduli and on higher-order
corrections that we do not discuss here.  
In analogy with the heterotic case, this result (\ref{e12}) is
interpreted as a mixing between
the twisted moduli $M_k$ and the untwisted ones $S,T_i$, which could in principle
be checked by an explicit string computation. 

The second, more involved, example, is the $Z_6'$ orientifold. Let us compare
again the string results obtained in section 5 and the effective
supergravity. Using (\ref{t060}) and (\ref{t8}) we find
\ba
{4 \pi^2 \over g_{SU(8)}^2} \!\!&=&\!\! {1 \over \ell} -2 s_2 m_2 \nonumber \\
&&\!\!\!\!-{1 \over 4} \left[ 24 {\rm Re} \ln \eta (U_2) + 4 \ln {\mu^2 \over M_I^2}
+ 6 \ln (\sqrt{G_2} {\rm Im}U_2 \mu^2) - 4 \ln (\sqrt{G_3} \mu^2) \right] \ ,
\nonumber \\
{4 \pi^2 \over g_{SU(4)}^2} \!\!&=&\!\! {1 \over \ell} + s_2 m_2 +s_1 m_1 
\nonumber \\
&&\!\!\!\!-{1 \over 4} \left[ -12 {\rm Re} \ln \eta (U_2) -2 \ln {\mu^2 \over M_I^2}
-3 \ln (\sqrt{G_2} {\rm Im}U_2 \mu^2) - 4 \ln (\sqrt{G_3} \mu^2) \right] 
\ , \label{e13}
\ea
where $s_1,s_2$ are numerical coefficients computed in (\ref{t060}). On
the other hand, the field theory relation (\ref{e4}) becomes   
\ba
{4 \pi^2 \over g_{SU(8)}^2} \!\!&=&\!\! {\rm Im}S -2 s_2 M_2 \nonumber \\
&&\!\!\!\!-{1 \over 4} \left[ 24 {\rm Re} \ln \eta (U_2) + 2 \ln (\sqrt{G_1} \mu^2) 
+ 6 \ln (\sqrt{G_2} {\rm Im}U_2 \mu^2) - 2 \ln (\sqrt{G_3} \mu^2) \right] \ ,
\nonumber \\
{4 \pi^2 \over g_{SU(4)}^2} \!\!&=&\!\! {\rm Im} S + s_2 M_2 +s_1 M_1 
\nonumber \\
&&\!\!\!\!-{1 \over 4} \left[ -12 {\rm Re} \ln \eta (U_2) - \ln (\sqrt{G_1} \mu^2) 
-3 \ln (\sqrt{G_2} {\rm Im}U_2 \mu^2) - 5 \ln (\sqrt{G_3} \mu^2) \right] 
\ . \label{e14}
\ea
Therefore the string theory and the field theory results are compatible provided
the following linear-chiral multiplet duality transformation is performed
\be
m_2 = {\rm Im}M_2 + {1 \over 4 s_2} \ln
(\sqrt{G_1 G_3} M_I^4) = {\rm Im}M_2 +
{1 \over 4 s_2} \ln ({{\rm Im}S \over {\rm Im}T_2}) \ . \label{e15}
\ee
Notice that in the $Z_3$ case where the moduli $M$ corresponds to ${\cal N}=1$ sectors
twisted with respect to the three complex planes, the linear-chiral multiplet 
duality (\ref{e12}) involves the full volume. 
In the $Z_6'$ case on the other hand, where $M_2$ corresponds to ${\cal N}=2$
sectors twisted with respect to the two-torii $T^1,T^3$ and untwisted with
respect to $T^2$, the linear-chiral multiplet duality (\ref{e15}) involves the volume of 
the twisted planes $T^1,T^3$.

Our results on threshold corrections can be used to discuss
heterotic - type I duality which was conjectured to hold for some
${\cal N}=1$ 4d vacua where perturbative spectra match, 
even though there is no regime where both sides are weakly coupled and there are no BPS
states to compare. The massless spectrum of the $Z_3$ case was described above
(heterotic side) and in section 4 (type I side). On the heterotic
side there is an anomalous gauge factor $U(1)_X$ \cite{ABPSS} which forces the twisted
matter fields 
${\bf (1,1)_{-4}}$ to get a VEV, breaking $U(1)_X$ and giving superpotential
masses to the $27$ twisted charged fields $ {\bf (1,8_s)_{+2}}$ \cite{K}. 
These VEV's blow-up the heterotic orbifold singularities and 
the resulting heterotic massless spectrum coincides with the one of type I 
at the orbifold point $m_k=0$. Notice that the $U(1)_X$ gauge field on the type
I side becomes massive without the need of any scalar VEV \cite{P}, and thus leaving
unbroken the global $U(1)_X$ symmetry which has a counter part on the heterotic 
side. After this blowing-up procedure, the heterotic threshold corrections are given
by the field theory expression (\ref{e9}, \ref{e10}), that does not depend 
on the heterotic string scale, as explained in section 7. 
Comparing this expression with the type I one-loop threshold corrections 
at the orbifold point $m_k=0$ (\ref{e011}) one finds a disagreement since the type I
string scale appears explicitly. A possible explanation could be the existence of
non-perturbative corrections, depending logarithmically on the string coupling, on one
of the two sides which is necessarily strongly coupled.
It is interesting to notice however that if the VEV of
the type I twisted moduli were vanish in the chiral basis $M_k=0$, the two one-loop
results would match. In view of the relation (\ref{e15}), this would imply that
the orbifold point should be unstable on the type I side as well, due to higher order
corrections, which should induce a Fayet-Iliopoulos (FI) term for the anomalous
$U(1)_X$  beyond one-loop, depending on the compactification radii. This
possibility seems though unlikely, in view of the arguments of ref. \cite{P}.

%%%%%%%%%%%%%%%%%%%%%%%%%%%%%%%%%%%%%%%%%%%%%%%%%%%%%%%%%%%%%%%%%%%%%%%%%%%%%%%%%%%%%%%%%%%
\section{Anomalous $U(1)$'s and gauge coupling unification}

It is interesting to discuss in more detail the anomaly cancellation
mechanism for the anomalous $U(1)$ factors in four-dimensional ${\cal N}=1$
orientifold models\footnote{More discussion can be found in the recent
papers \cite{LLN}.}. Let us start for simplicity with the $Z_3$ example
discussed in section 4, where a linear symmetric combination $M$ of the
$27$ twisted moduli couples to the gauge fields 
\be
f_a = S + s_a M \ , \label{g1}
\ee 
and the coefficients $s_a$ were computed in (\ref{d18}). The model
contains a single anomalous $U(1)_X$ with the gauge generator
$Q_X=(1^{12},0^4)$ in a complex $U(16)$ basis. Under a $U(1)_X$ gauge
transformation with (superfield) parameter $\Lambda$, there are
cubic gauge anomalies. The generalized Green-Schwarz mechanism requires a shift
of the twisted moduli field combination $M$
\be
V_X \rightarrow V_X + {i \over 2} (\Lambda-{\bar \Lambda}) \ , \ M
\rightarrow M + {1 \over 2} \ \epsilon \ \Lambda \ , \label{g2}
\ee
such that the gauge-invariant combination appearing in the K{\"a}hler
potential is $i(M-{\bar M})-\epsilon V_X$. The mixed anomalies are
cancelled provided the following condition holds
\be
{\epsilon \over 4 \pi^2} = {C_{SU(12)} \over s_{SU(12)}}= {C_{SO(8)} \over s_{SO(8)}}=
{C_{U(1)_X} \over s_{U(1)_X}} \ . \label{g3}
\ee
The value of $\e$ was computed in (\ref{d21})
\be
\e =  \sqrt{2 \over N \pi^3} \sum_k \prod_{i=1}^3 | \sin \pi k v_i|^{1 \over 2}
(-i {\rm tr} Q_X \g^k) \  \label{g30}
\ee
and $(-i {\rm tr} Q_X \g)=12 \sqrt{3}$.
By using the values of the cubic anomalies $(C_{SU(12)},C_{SO(8)},C_{U(1)})$
$=$ $(1/4 \!\pi^2) \!(-18,36,-432)$ and (\ref{d18}) it is
straightforward to check (\ref{g3}), which is the direct check of the
anomaly cancellation mechanism.
By supersymmetry arguments, we can also write down the D-terms which
encode the induced Fayet-Iliopoulos term
\be
V_D = {g^2 \over 2} (\sum_A X_A K_A \Phi^A + \epsilon K_M M_P^2 )^2 \ , \label{g4}
\ee
where $\Phi^A$ denotes the set of charged chiral fields of $U(1)_X$
charge $X_A$, $K_A=\partial K / \partial \Phi^A$ and analogously for
$K_M$.

The above discussion generalizes easily in the case of more anomalous
$U(1)_\a$ (${\a}=1 \cdots N_X)$ and more linear combinations of twisted
moduli fields $M_k$ coupling to gauge fields. In this case (\ref{g1}) becomes
\be
f_a = S + \sum_k s_{ak} M_k \ , \label{g5}
\ee
and (\ref{g2}) generalizes to 
\be
V_\a \rightarrow V_\a + {i \over 2} (\Lambda_\a -{\bar \Lambda}_\a) \ , \ M_k
\rightarrow M_k + {1 \over 2} \ \epsilon_{k\a} \ \Lambda_\a \ , \label{g6}
\ee
in an obvious notation. Cancelation of gauge anomalies ${\rm tr} X_\a Q_a^2$
described by the coefficients $C_{{\a}a}$ ask for the Green-Schwarz conditions
\be
C_{{\a}a} = {1 \over 4 \pi^2} \sum_k s_{ak} \epsilon_{k\a} \ , \label{g7}
\ee
valid for each ${\a},a$. The gauge-invariant field combination appearing in the
K{\"a}hler  potential is $ i(M_k-{\bar M}_k)- \sum_\a \epsilon_{k\a} V_\a$ and
generates, by supersymmetry, the D-terms
\be
V_D = \sum_\a {g_{\a}^2 \over 2} (\sum_A X_A^\a K_A \Phi^A + \sum_k
\epsilon_{k\a} 
{\partial K \over \partial M_k} M_P^2 )^2 \ . \label{g8}
\ee

In the above discussion, we neglected the additional complication of linear versus
chiral multiplet that arises from the change of basis of the type (\ref{e12}).
Although a detailed analysis is needed to be done in the presence of several linear
multiplets, it appears that the gauge invariant combination entering in the K{\"a}hler 
potential involves the scalar of the linear multiplet $m_k$ instead of the chiral one
$M_k$, as in the expression of gauge couplings. We will now show that, at least in the
examples studied in this work, the linear combinations of twisted moduli appearing in
gauge couplings are the same with the combinations entering in the anomalous $U(1)$
D-terms. Therefore, the vanishing of the latter at the point with maximal gauge symmetry
determines the VEVs of the corresponding blowing-up directions and removes the twisted
moduli dependence of gauge couplings.

In fact, in the $Z_3$ case there is one anomalous $U(1)$ with the corresponding FI term
proportional to the symmetric combination of the 27 blowing-up modes, which
also appears in the expression of gauge couplings, as discussed in section 4. Requiring
that the non-abelian gauge group remains unbroken, the vanishing of the FI-term fixes the
symmetric linear combination of the twisted moduli, removing the arbitrariness in the
gauge couplings. 
At the one-loop level, this selects the orbifold point $m_k=0$ \cite{P}
(or equivalently Im$M_k={1\over 12c}\ln (VM_I^6)$), 
implying that physical gauge couplings are  moduli
independent (up to one loop) and unify at the string scale. On the other hand,
if there are higher order corrections that destabilize the orbifold vacuum
and fix the twisted moduli VEVs at the point $M_k=0$ as discussed in the previous
section, the unification scale would be determined by the size of the compact space.
This is an open important question that deserves further investigation.\footnote{ In
ref. \cite{I} it was assumed that the twisted moduli appearing in
the gauge coupling are in the chiral basis $M_k$, with a non-vanishing
VEV, leading to a logarithmic volume dependence in the gauge couplings.} 
In the $Z_6'$ orientifold,
there are two linear combinations of twisted moduli fields entering into
the expression of gauge couplings (\ref{u12}), (\ref{t060}). On the other hand,
the model has two anomalous U(1)'s in each of the D9 and D5 brane sectors.
A simple inspection shows that the vanishing of the corresponding FI terms 
(without breaking the non-abelian gauge symmetry)
fixes both combinations appearing in the gauge couplings, removing again the
arbitrariness. 
At the one-loop level, this selects as before the orbifold point $m_k=0$.
In this example, the ${\cal N}=1$ sectors contribute to the running up to
the type I string scale, while the ${\cal N}=2$ sectors lead to threshold
corrections depending on two compact tori, associated to the two ${\cal N}=2$ 
sectors of the model ($\theta^2$ and $\theta^3$). The issue of higher loop
corrections is similar to the previous ($Z_3$) example.

The situation simplifies in the case of freely acting type I
orbifold compactifications \cite{ADS,ADDS}. The couplings of the gauge fields to the
twisted  moduli and the presence of anomalous $U(1)$'s
is determined by the (large or small radius) limit where supersymmetry is restored.
For instance, in all the examples studied in \cite{ADS,ADDS}, in particular in the
ones discussed in section 6, there are no anomalous
$U(1)$'s neither couplings to twisted moduli.

%%%%%%%%%%%%%%%%%%%%%%%%%%%%%%%%%%%%%%%%%%%%%%%%%%%%%%%%%%%%%%%%%%%%%%%%%%%%%%%%%%%%%%%
\section{\bf Conclusions and discussions}

The primary goal of this paper was the study of threshold corrections to
the gauge couplings in four-dimensional type I orientifolds. The method
we use, developed in \cite{BF}, \cite{B} consists in coupling a background magnetic
field $B$ to the Chan-Paton charges of the open strings and computing
the quadratic terms $B^2$ in a weak-field expansion. We find that, in
the ${\cal N}=2$ sectors of the orientifolds the string oscillators
decouple and the result is entirely due to Kaluza-Klein modes of the
complex two-tori. This is in agreement with the expectation that only
BPS states contribute to the threshold corrections in these sectors \cite{DL,BF,B}.  
For a rectangular torus of radii $R_1,R_2$ these corrections are proportional
to $\ln (R_1 R_2) + f(R_1/R_2)$, where the function $f$ diverges linearly $f \sim R_1/R_2$ 
in the $R_1 >> R_2$  limit. For phenomenological purposes, the linear term
can be used in the accelerated unification scenario of \cite{DDG}, while
the logarithmic correction can acomodate for a more traditional unification \cite{B}. 
In the ${\cal N}=1$ sectors the string oscillators do not decouple and
the corrections are independent of the compact volume in the orbifold
limit. By identifying the string IR divergences with the effective theory running, a
string formula for the one-loop beta coefficients of the effective field theory
was derived in (\ref{t4}). We showed in section 6 that the dependence on the compact
radii in models where supersymmetry is spontaneously broken by
compactification \cite{ADS,ADDS} is milder. In particular, in the Scherk-Schwarz model where
the branes are parallel to the coordinate used to break supersymmetry $R_1 \rightarrow \infty$,
the linear behaviour is absent. On the other hand, in the M-theory model
where  the branes are orthogonal to the breaking
coordinate $R_1' \rightarrow \infty$ (where $R_1' = 1/R_1M_I^2$ is the
T-dual coordinate) the logarithmic corrections disappear as well and the thresholds are 
exponentially suppressed $e^{-R_1'R_2M_I^2}$ in the large radius limit, fact that could have 
interesting phenomenological implications.
  
We explicitly computed the UV divergences in $B^4$ in the one-loop open
string amplitudes, interpreted
in the closed string channel as the propagation of the massless twisted
moduli $M_k$ coupling (at the disk level) to the gauge fields. By
comparison of the two pictures, we derived the explicit form of these couplings
(\ref{u12}), (\ref{d18}), (\ref{t060}), first discussed in
\cite{IRU}. It turns out that the twisted moduli of all sectors (except
the sector $\theta^{N/2}$ for $N$ even) generically couple to the gauge
fields, fact that was also justified by supersymmetry arguments in
section 7. Similarly we can couple a background magnetic field $B'$ to the
anomalous $U(1)$ factors. In this case, the $B'^2$ UV divergences in the
open sector amplitudes allowed us to single out the mixing of the RR
twisted moduli with the anomalous gauge fields (\ref{d21}). Using
these results, we discussed the Fayet-Iliopoulos terms for the anomalous
U(1) factors in section 8, as well as the generalized Green-Schwarz mechanism
\cite{S} in section 7. 

The method we used to obtain the above results for the D9 brane gauge groups can be
applied  in a straightforward way to the D5 branes gauge groups, as well, by coupling a
background magnetic field to the Dirichlet strings. The difference
compared to D9 branes is that the threshold corrections can depend only 
on the compact torus contained inside the D5 branes world-volume, instead of the
three torii available for the D9 branes. Moreover, the couplings (\ref{t060}) of the
twisted moduli to the D5 gauge fields exist only for D5 branes located
at orbifold fixed points and are nonvanishing only for
twisted moduli living in the fixed point where the D5 brane is located.

We also performed a comparison between the one-loop corrected string gauge couplings
and the general field theory results \cite{KL1,KL2}. This was done explicitly
in the examples of $Z_3$ and $Z_6'$ orientifolds. We found that the two results differ
by gauge group dependent corrections, unlike the heterotic orbifolds, where the
difference is given by a universal term. In the latter case, this difference is
explained by a chiral-linear multiplet duality (\ref{e11}) for the dilaton multiplet,
which involves the Green-Schwarz term related to the one-loop correction to the
K{\"a}hler potential. In type I orientifolds, the compatibility between the two
results ask for specific chiral-linear multiplet
duality relations (\ref{e12}), (\ref{e15}) for the twisted moduli, which
are described by linear (chiral) multiplets $m_k$ ($M_k$) in the string (field theory)
basis. This amounts to gauge group dependent corrections to the gauge couplings
due to the non-universal tree-level (disk) couplings of the twisted moduli to the
gauge fields.
In analogy with the heterotic case, this requires loop corrections to the twisted
moduli K{\"a}hler potential which would be interesting to be explicitly computed.

Our results on threshold corrections were used to discuss in section 7
heterotic - type I duality which was conjectured to hold for the $Z_3$ 4d vacuum. 
Comparing the threshold
corrections of the two sides we find a disagreement which raises doubts on the
perturbative validity of the conjectured duality.  

We finish by discussing the implication of our results on the unification of
gauge couplings. At the level of ${\cal N}=2$ compactifications, in all orientifolds
with the exception of $Z_2$, the tree-level (disk) gauge couplings depend linearly on
the VEVs of twisted moduli which correspond to exact flat directions of the
scalar  potential. As a result, unification is in general lost, although it is preserved
in special examples  (such as $Z_3$ in 6d), where the coefficients of these couplings
are proportional to the (4d) beta-functions. In the latter case, however, the
unification scale is an arbitrary parameter, depending on the VEVs of the twisted
moduli. 

For ${\cal N}=1$ orientifolds, the generic presence of anomalous $U(1)$'s fixes 
the VEVs of the linear combinations of the twisted moduli that appear in gauge
couplings, at least in the examples we discussed in this work. At the point of maximal
gauge symmetry, the dependence on twisted moduli of gauge couplings vanish (up to
one-loop order), and one is left with the dependence on geometric moduli coming from the 
${\cal N}=2$ sectors. As a result, generically  ${\cal N}=1$ sectors contribute to the
running up to the type I string scale, while the ${\cal N}=2$ sectors run up to the
corresponding compactification radii. For low scale string models, our results imply
that the only way to achieve gauge coupling unification is by using the running
controled by the ${\cal N}=2$ beta-functions, that can be either power-like or
logarithmic.

%%%%%%%%%%%%%%%%%%%%%%%%%%%%%%%%%%%%%%%%%%%%%%%%%%%%%%%%%%%%%%%%%%%%%%%%%%%%%%%%%%%%%%%%%%%%%%%%%
\noindent{\bf Acknowledgements} 

We would like to thank K. Benakli, J.-P. Derendinger, E. Kiritsis, C. Kounnas,
S. Lavignac, H.P. Nilles, A. Sagnotti, S. Stieberger and  A. Uranga for discussions.

\vfill
\eject
%%%%%%%%%%%%%%%%%%%%%%%%%%%%%%%%%%%%%%%%%%%%%%%%%%%%%%%%%%%%%%%%%%%%%%%%%%%%%%%%%%%%%%%%
\setcounter{section}{0}   %  starts Appendix lettering at "A"
\Appendix{}

  For the reader's convenience we collect in this apendix the
  definitions, transformation properties and some identities among the
  modular functions that are used in the text. For a more
  extensive list see for instance \cite{E}. The Dedekind function is
  defined by the usual product formula (with $q=e^{2\pi i\tau}$)
\be
\eta(\tau) = q^{1\over 24} \prod_{n=1}^\infty (1-q^n)\ .
\ee
The Jacobi $\vartheta$-functions with general characteristic and
  arguments  are
\be
\vartheta [{\a \atop \b }] (z\vert\tau) = \sum_{n\in Z}
e^{i\pi\tau(n-\a)^2} e^{2\pi i(z- \b)(n-\a)}\ .
\ee
We give also the product formulae for the four special $\vartheta$-functions
\be\eqalign{
\ \ \vartheta_1(z\vert\tau) &\equiv \vartheta \left[{{1\over 2} \atop {1\over 2} }\right]
  (z\vert\tau) = 2q^{1/8}{\rm sin}\pi z\prod_{n=1}^\infty
  (1-q^n)(1-q^ne^{2\pi i z})(1-q^ne^{-2\pi i z})\cr
\ \ \vartheta_2(z\vert\tau) &\equiv \vartheta \left[{{1\over 2} \atop 0 }\right]
  (z\vert\tau) = 2q^{1/8}{\rm cos}\pi z\prod_{n=1}^\infty
  (1-q^n)(1+q^ne^{2\pi i z})(1+q^ne^{-2\pi i z})\cr
\ \ \vartheta_3(z\vert\tau) &\equiv \vartheta \left[{0 \atop 0 }\right]
  (z\vert\tau) = \prod_{n=1}^\infty
  (1-q^n)(1+q^{n-1/2}e^{2\pi i z})(1+q^{n-1/2}e^{-2\pi i z})\cr
\ \ \vartheta_4(z\vert\tau) &\equiv \vartheta \left[{0 \atop {1\over 2} }\right]
  (z\vert\tau) = \prod_{n=1}^\infty
  (1-q^n)(1-q^{n-1/2}e^{2\pi i z})(1-q^{n-1/2}e^{-2\pi i z})\cr\ .
}
\ee
The $\vartheta_a$ for $a=2,3,4$ are even functions of $z$, while
  $\vartheta_1$ is an odd function whose first derivative at zero
is
\be
\vartheta_1^\prime(0) = 2\pi\eta^3\ .
\ee
The modular properties of these functions are described by
\be
\eta(\tau+1) = e^{i\pi/12}\eta(\tau)\ \ , \ \
\vartheta \left[{\a \atop {\b}}\right] \left({z} \Bigl|
  {\tau+1}\right)=
e^{-i\pi\a(\a-1)}\vartheta 
\left[{\a \atop {\a+\b-{1\over 2}}}\right] \left({z} \Bigl|
  {\tau}\right)
\ee
\be
\eta(-1/\tau) = \sqrt{-i\tau}\; \eta(\tau)\ \ , \ \ 
\vartheta \left[{\a \atop {\b}}\right] \left({z \over \tau} \Bigl| {-1 \over \tau}\right)=
\sqrt{-i \tau} \ e^{2 i \pi \a \b +{i \pi z^2 / \tau}} \ 
\vartheta \left[{{\b} \atop - \a}\right] (z | \tau ) \ ,  \label{f8}
\ee

A very  useful  identity is 
\be\eqalign{
 &\sum_{\a,\b=0,1/2}  \eta_{\a,\b} \
\vartheta \left[{\a \atop \b }\right] (z) \ \prod_{i=1}^3  \
  \vartheta\left[{\a \atop {\b + v_i}}\right] = 
\cr
- 2\; \vartheta_1
\left(-{z \over 2}\right)  & \vartheta_1
 \left({z  -v_1+v_2+v_3 
\over 2}\right)\;
 \vartheta_1 \left({z +v_1-v_2+v_3 \over 2}\right)\; \vartheta_1 
\left({z +v_1+v_2 -v_3 \over 2}\right)\;
\ , \cr} 
\label{a1}
\ee
valid for $v_1+v_2+v_3=0$. By  taking the second derivative of
(\ref{a1})  at zero argument
 it is easy to prove the  following  identities
\ba
&& \sum_{\a,\b=0,1/2}  (-1)^{2 \a } \eta_{\a,\b} \
\frac{\vartheta^{''}[{\a \atop \b }] \vartheta [{\a \atop \b }]}{\eta^6} \  
\frac{\vartheta[{\a \atop {\b + kv_i}}] \vartheta[{\a \atop {\b + kv_j}}]}
{\vartheta[{1/2 \atop {1/2 + kv_i}}] \vartheta[{1/2 \atop {1/2 + kv_j}}]}  
= -4 \pi^2 \ , \ k(v_i+v_j) = 1 (mod \ 2) \ , \nonumber \\
&& \sum_{\a,\b=0,1/2} \eta_{\a,\b} \
\frac{\vartheta^{''}[{\a \atop \b }] \vartheta [{\a \atop \b }]}{\eta^6} \  
\frac{\vartheta[{\a +1/2 \atop {\b + kv}}] \vartheta[{\a +1/2 \atop {\b - kv}}]}
{\vartheta[{0 \atop {1/2 + kv}}] \vartheta[{0 \atop {1/2 - kv}}]}  
= 4 \pi^2 \  \ , \nonumber \\
&& \sum_{\a,\b=0,1/2}  (-1)^{2 \a } \eta_{\a,\b} \
\frac{\vartheta^{''}[{\a \atop \b }] \vartheta [{\a \atop \b }]}{\eta^6} \  
\frac{\vartheta^2 [{\a +1/2 \atop \b }]}
{\vartheta^2 [{0 \atop 1/2 }]}  
= -4 \pi^2 \ , \nonumber \\
&& \sum_{\a,\b=0,1/2} \eta_{\a,\b} \
\frac{\vartheta^{''}[{\a \atop \b }] \vartheta [{\a \atop \b }]}{\eta^6} \  
\frac{\vartheta[{\a \atop {\b + 1/2}}] \vartheta[{\a \atop {\b - 1/2}}]}
{\vartheta[{1/2 \atop 0}]}  = 4 \pi^2 \ , \label{a2}
\ea
which help us to prove that the oscillator contributions to the threshold corrections 
decouple for the ${\cal N}=2$ sectors $k=2,3,4$ of the $Z_6'$ orientifold and in general 
for ${\cal N}=2$ sectors of any four-dimensional type I orientifold.

In the contributions from the ${\cal N}=1$ sectors coming from
${\cal A}_{95}$ in section 5, we used the
following modular identity (also valid for $v_1+v_2+v_3=0$)
\ba
\sum_{\a,\b=0,1/2} \!\!\! \eta_{\a,\b} 
\frac{\vartheta^{''}[{\a \atop \b }]}{\eta^3}  
\frac{\vartheta[{\a \atop {\b + kv_3}}]}{\vartheta[{1/2 \atop {1/2 + kv_3}}]} 
\prod_{i=1}^2   
\frac{\vartheta[{\a \atop {\b + kv_i}}]}{\vartheta[{1/2 \atop {1/2 + kv_i}}]} 
\!=\! -2 \pi \left \{ \frac{\vartheta^{'} [{1/2 \atop {1/2 - kv_3}}]}
{\vartheta [{1/2 \atop {1/2 - kv_3}}]} + \sum_{i=1}^2 \frac{\vartheta^{'} 
[{0 \atop {1/2 - kv_i}}]}{\vartheta [{0 \atop {1/2 - kv_i}}]} \right \} \ . 
\label{a3}
\ea
In taking the infrared limit for the $Z_6'$ model, we need also the formula
\be
\lim_{q \rightarrow 0} \frac{\vartheta^{'} 
[{0 \atop {1/2 - kv_i}}]}{\vartheta [{0 \atop {1/2 - kv_i}}]} =0 \ . \label{a4}
\ee
%%%%%%%%%%%%%%%%%%%%%%%%%%%%%%%%%%%%%%%%%%%%%%%%%%%%%%%%%%%%%%%%%%%%%%%%%%%%%%%%%%%%%%
% \setcounter{section}{0}   %  starts Appendix lettering at "A"
\Appendix{$Z_6'$ one-loop amplitudes}

The annulus amplitudes (for vanishing magnetic field) relevant for the computation of
section 5 of the $Z_6'$ model are
\ba
{\cal A}_{99}^{(k)} \!\!&\!\!\!=\!\!\!&\! {1 \over 8 \pi^4 t^2}\!\!
\sum_{\a,\b=0,1/2}  \eta_{\a,\b} \
\frac{\vartheta[{\a \atop \b}]}{\eta^3} \ \prod_{i=1}^3 (-2\sin \pi k v_i)
\  \frac{\vartheta[{\a \atop {\b + kv_i}}]}{\vartheta[{1/2 \atop {1/2 +kv_i}}]} \,  \
( {\rm tr} {\g}_9^{k})^2 \ , \ k=1,5 \nonumber \\
{\cal A}_{99}^{(k)} \!\!&\!\!\!=\!\!\!&\! {1 \over 8 \pi^4 t^2}\!\! \sum_{\a,\b=0,1/2}  \eta_{\a,\b} \
\frac{\vartheta[{\a \atop \b}] \vartheta[{\a \atop {\b +kv_2}}]}{\eta^6} \ \prod_{i=1,3} 
(2\sin \pi k v_i)
\  \frac{\vartheta[{\a \atop {\b + kv_i}}]}{\vartheta[{1/2 \atop {1/2 +kv_i}}]} \,  \
( {\rm tr} {\g}_9^{k})^2 \ \Gamma_2^{(2)} \ , \ k= 2,4 \nonumber \\
{\cal A}_{99}^{(3)} \!\!&\!\!\!=\!\!\!& {1 \over 8 \pi^4 t^2}\!\!
\sum_{\a,\b=0,1/2}  \eta_{\a,\b} \
\frac{\vartheta[{\a \atop \b}] \vartheta[{\a \atop {\b +3v_3}}]}{\eta^6} \ \prod_{i=1,2} 
(2\sin \pi 3 v_i)
\  \frac{\vartheta[{\a \atop {\b + 3v_i}}]}{\vartheta[{1/2 \atop {1/2 +3v_i}}]} \,  \
( {\rm tr} {\g}_9^{3})^2 \ \Gamma_3^{(2)} \ , \label{a10}
\ea
for the NN (99) sector. The annulus amplitudes from the ND (95) sector are
\ba
{\cal A}_{95}^{(k)} \!\!&\!\!\!=\!\!\!&\! {-1 \over 2 \pi^4 t^2}\!\!
\sum_{\a,\b=0,1/2} \!\! \eta_{\a,\b} \,
\frac{\vartheta[{\a \atop \b}]}{\eta^3}
\frac{ \sin \pi k v_3 \vartheta[{\a \atop {\b + kv_3}}]}
{\vartheta[{1/2 \atop {1/2 +kv_3}}]} \,
\prod_{i=1}^2
\  \frac{\vartheta[{{\a+1/2} \atop {\b + kv_i}}]}
{\vartheta[{0 \atop {1/2 +kv_i}}]} \,
{\rm tr} \g_9^k \, {\rm tr} \g_5^k \ , \ k\!\!=\!\!1,\!2,\!4,\!5\! \nonumber \\
{\cal A}_{95}^{(k)} \!\!&\!\!\!=\!\!\!&\!\! {1 \over 4 \pi^4 t^2}\!
\sum_{\a,\b=0,1/2}  \eta_{\a,\b} \,
\frac{\vartheta[{\a \atop \b}] \vartheta[{\a \atop \b +kv_3}] }{\eta^6} \,
\prod_{i=1}^2
\  \frac{\vartheta[{{\a+1/2} \atop {\b + kv_i}}]}
{\vartheta[{0 \atop {1/2 +kv_i}}]} \,
{\rm tr} \g_9^k \, {\rm tr} \g_5^k \ \Gamma_3^{(2)} \ , \ k=0,3 \ . \label{a11}
\ea
Similarly, the supersymmetric M{\"o}bius amplitudes from the 99 sector are
\ba
{\cal M}_9^{(k)} \!\!&\!\!=\!\!&\! -{1 \over 8 \pi^4 t^2}\! \!\!
\sum_{\a,\b=0,1/2} \!\! \eta_{\a,\b} \
\frac{\vartheta[{\a \atop \b}]}{\eta^3} \ \prod_{i=1}^3 (-2\sin \pi k v_i)
\  \frac{\vartheta[{\a \atop {\b + kv_i}}]}{\vartheta[{1/2 \atop {1/2 +kv_i}}]} \,  \
{\rm tr} {\g}_9^{2k} \ , \ k=1,5 \nonumber \\
{\cal M}_9^{(k)} \!\!&\!\!=\!\!&\! -{1 \over 8 \pi^4 t^2}\!\!\!
\sum_{\a,\b=0,1/2} \!\! \eta_{\a,\b} \
\frac{\vartheta[{\a \atop \b}] \vartheta[{\a \atop {\b +kv_2}}]}{\eta^6} \ \prod_{i=1,3} 
(2\sin \pi k v_i)
\  \frac{\vartheta[{\a \atop {\b + kv_i}}]}{\vartheta[{1/2 \atop {1/2 +kv_i}}]} \,  \
{\rm tr} {\g}_9^{2k} \ \Gamma_2^{(2)} \ , \ k= 2,4 \nonumber \\
{\cal M}_9^{(3)} \!\!&\!\!=\!\!&\! -{1 \over 8 \pi^4 t^2}\!\!\!
\sum_{\a,\b=0,1/2} \!\! \eta_{\a,\b} \
\frac{\vartheta[{\a \atop \b}] \vartheta[{\a \atop {\b +3v_3}}]}{\eta^6} \ \prod_{i=1,2} 
(2\sin \pi 3 v_i)
\  \frac{\vartheta[{\a \atop {\b + 3v_i}}]}{\vartheta[{1/2 \atop {1/2 +3v_i}}]} \,  \
{\rm tr} {\g}_9^{6} \ \Gamma_3^{(2)} \ . \label{a12}
\ea
In the presence of background magnetic field coupled to the D9 brane gauge group,
the above annulus 99 amplitudes become
\ba
{\cal A}_{99}^{(k)} \!\!&\!\!\!=\!\!&\!\!\! {-iB \over \pi^3 t} \! {\rm tr}
\left( (Q \g_9^k \! \otimes \g_9^k \!+\! \g_9^k \otimes Q \g_9^k)  \!\!\!\!
\sum_{\a,\b=0,1/2} \!\!\!\!  \eta_{\a,\b}
\frac{\vartheta[{\a \atop \b }]({i \e t \over 2})}{\vartheta[{1/2 \atop
1/2}] ({i \e t \over 2})} \right) \!\!
\prod_{i=1}^3 \!\!\sin \pi k v_i
\  \frac{\vartheta[{\a \atop {\b + kv_i}}]}{\vartheta[{1/2 \atop 
{1/2 + kv_i}}]}  \  \label{a13}
\ea
for the ${\cal N}=1$ sectors $k=1,5$,
\ba
{}\!\!\!\!\!{\cal A}_{99}^{(k)} \!\!&\!\!\!=\!\!&\!\!\! {iB \over 2 \pi^3 t}
\! {\rm tr}
\left( (Q \g_9^k \otimes \g_9^k \!+\! \g_9^k \otimes Q \g_9^k)
 \!\!\!\!\!\!
\sum_{\a,\b=0,1/2} \!\!\!\!\!\!  \eta_{\a,\b} \
\frac{\vartheta[{\a \atop \b }]({i \e t \over 2}) \vartheta[{\a \atop \b +kv_2}]}
{\vartheta[{1/2 \atop 1/2 }]({i \e t \over 2}) \eta^3} \right) \!\!\!
\prod_{i=1,3} \! \sin \pi k v_i
\frac{\vartheta[{\a \atop {\b + kv_i}}]}{\vartheta[{1/2 \atop 
{1/2 + kv_i}}]} \Gamma_2^{(2)} \  \nonumber
\ea
for the ${\cal N}=2$ sectors $k=2,4$ and
\ba
{}\!\!\!\!\!{\cal A}_{99}^{(3)} \!\!&\!\!\!=\!\!&\!\!\! {iB \over 2 \pi^3 t}
\!\ {\rm tr}
\left( (Q \g_9^3 \otimes \! \g_9^3 \!+\! \g_9^3 \otimes \! Q \g_9^3)
 \!\!\!\!\!
\sum_{\a,\b=0,1/2} \!\!\!\!\!  \eta_{\a,\b} \!\!
\frac{\vartheta[{\a \atop \b }]({i \e t \over 2}) \vartheta[{\a \atop \b +3v_3}]}
{\vartheta[{1/2 \atop 1/2 }]({i \e t \over 2}) \eta^3} \right) \!\!
\prod_{i=1,2} \! \sin \pi 3 v_i
\  \frac{\vartheta[{\a \atop {\b + 3v_i}}]}{\vartheta[{1/2 \atop 
{1/2 + 3v_i}}]} \Gamma_3^{(2)} \nonumber
\ea
for the ${\cal N}=2$ sector $k=3$. The annulus 95 amplitudes in the presence of the magnetic
field become
\ba
{\cal A}_{95}^{(k)} \!\!&\!\!\!=\!\!&\!\!\!  {-iB \over 2 \pi^3 t} 
{\rm tr} \left( Q \g_9^k \otimes \g_5^k 
 \!\!\!\!
\sum_{\a,\b=0,1/2} \!\!\!\!  \eta_{\a,\b} \!
\frac{\vartheta[{\a \atop \b }]({i \e t \over 2})}
{\vartheta[{1/2 \atop 1/2 }]({i \e t \over 2})} \right) \!
\frac{ \!\!\sin \pi k v_3 \vartheta[{\a \atop {\b + kv_3}}]}{\vartheta[{1/2 \atop 
{1/2 + kv_3}}]}
\prod_{i=1,2} \frac{\vartheta[{\a +1/2 \atop {\b + kv_i}}]}{\vartheta[{0 \atop 
{1/2 + kv_i}}]}  \  \label{a16}
\ea
for $k=1,2,4,5$ and
\ba
{\cal A}_{95}^{(k)} \!\!&\!\!\!=\!\!&\!\!\! {iB \over 4 \pi^3 t} 
{\rm tr} \left( Q \g_9^k \otimes \g_5^k
 \!\!\!\!
\sum_{\a,\b=0,1/2} \!\!\!\!  \eta_{\a,\b} \
\frac{\vartheta[{\a \atop \b }]({i \e t \over 2}) \vartheta[{\a \atop {\b + kv_3}}]}
{\vartheta[{1/2 \atop 1/2 }]({i \e t \over 2}) \eta^3} \right) \!
\ \prod_{i=1,2} \  \frac{\vartheta[{\a +1/2 \atop {\b + kv_i}}]}{\vartheta[{0 \atop 
{1/2 + kv_i}}]} \ \Gamma_3^{(2)}  \  \label{a17}
\ea
for $k=0,3$. Similarly, the M{\"o}bius amplitudes in the Neumann case become
\ba
{\cal M}_9^{(k)} \!\!&\!\!\!=\!\!\!&\!\!\! {2iB \over \pi^3 t} {\rm tr}
\left( Q \g_9^{2k} \!\!\!\!
\sum_{\a,\b=0,1/2}  \eta_{\a,\b} \
\frac{\vartheta[{\a \atop \b }]({i \e t \over 2})}{\vartheta[{1/2 \atop
1/2 }]({i \e t \over 2})} \right) 
\prod_{i=1}^3 \!\! \sin \pi k v_i
\  \frac{\vartheta[{\a \atop {\b + kv_i}}]}{\vartheta[{1/2 \atop {1/2 + kv_i}}]}  
 \ , \label{a18}
\ea
from the ${\cal N}=1$ sectors $k=1,5$,
\ba
{\cal M}_9^{(k)} \!\!&\!\!\!=\!\!&\!\!\!- {iB \over  \pi^3 t}  
{\rm tr} \left( Q \g_9^{2k} 
 \!\!\!\!
\sum_{\a,\b=0,1/2} \!\!\!\!  \eta_{\a,\b}
\frac{\vartheta[{\a \atop \b }]({i \e t \over 2}) \vartheta[{\a \atop \b +kv_2}]}
{\vartheta[{1/2 \atop 1/2 }]({ i \e t \over 2}) \eta^3} \right) \!\!
\prod_{i=1,3} \! \sin \pi k v_i
\  \frac{\vartheta[{\a \atop {\b + kv_i}}]}{\vartheta[{1/2 \atop 
{1/2 + kv_i}}]} \Gamma_2^{(2)} \  \label{a19}
\ea
from the ${\cal N}=2$ sectors $k=2,4$ and
\ba
{\cal M}_9^{(3)} \!\!&\!\!\!=\!\!&\!\!\!- {iB \over \pi^3 t} 
{\rm tr} \left( Q \g_9^{6} 
 \!\!\!\!
\sum_{\a,\b=0,1/2} \!\!\!\!  \eta_{\a,\b}
\frac{\vartheta[{\a \atop \b }]({i \e t \over 2}) \vartheta[{\a \atop \b +3v_2}]}
{\vartheta[{1/2 \atop 1/2 }]({ i \e t \over 2}) \eta^3} \right) \!\!
\prod_{i=1,2} \! \sin \pi 3 v_i
\  \frac{\vartheta[{\a \atop {\b + 3v_i}}]}{\vartheta[{1/2 \atop 
{1/2 + 3v_i}}]} \Gamma_3^{(2)} \ 
\  \label{a20}
\ea
from the ${\cal N}=2$ sector $k=3$.
 
We also display here the various amplitudes in the transverse channel, necessary in
order to investigate the UV behaviour. The annulus amplitudes are
\ba
{1 \over l} {\cal A}_{99}^{(k)} \!\!&\!\!\!=\!\!&\!\!\! {-iB \over \pi^3} \! 
 {\rm tr}
\left( (Q \g_9^k \otimes \! \g_9^k \!+\! \g_9^k \otimes \! Q \g_9^k)
 \!\!\!\!
\sum_{\a,\b=0,1/2} \!\!\!\!  \eta_{\a,\b}
\frac{\vartheta[{\a \atop \b }](\e)}{\vartheta[{1/2 \atop 1/2 }](\e)} \right) \!
\prod_{i=1}^3 \!\! \sin \pi k v_i
\  \frac{\vartheta[{\a +kv_i \atop \b }]}{\vartheta[{1/2 +kv_i \atop 1/2}]}  \  
\label{a21}
\ea
for the ${\cal N}=1$ $k=1,5$ sectors,
\ba
{}\!\!\!\!\!{1 \over l} {\cal A}_{99}^{(k)} \!\!&\!\!\!=\!\!&\!\!\! {B v_2\over
4 \pi^3} 
 {\rm tr} \left( (Q \g_9^k \! \otimes \! \g_9^k \!+\! \g_9^k \otimes \! Q \g_9^k)
 \!\!\!\!\!\!
\sum_{\a,\b=0,1/2} \!\!\!\!\!\!  \eta_{\a,\b} \
\frac{\vartheta[{\a \atop \b }](\e) \vartheta[{{\a +kv_2} \atop \b }]}
{\vartheta[{1/2 \atop 1/2 }](\e) \eta^3} \right) \!\!\!
\prod_{i=1,3} \! \sin \pi k v_i
\frac{\vartheta[{\a +kv_i \atop \b }]}{\vartheta[{1/2 +kv_i \atop 
 1/2}]} W_2^{(2)} \nonumber
\ea
for the ${\cal N}=2$ sectors $k=2,4$ and
\ba
{}\!\!\!\!\!{1 \over l} {\cal A}_{99}^{(3)} \!\!&\!\!\!=\!\!&\!\!\! {B v_3\over
4 \pi^3} 
 {\rm tr} \left( (Q \g_9^3 \otimes \g_9^3 \!+\! \g_9^3 \otimes Q \g_9^3)
 \!\!\!\!\!\!
\sum_{\a,\b=0,1/2} \!\!\!\!\!\!  \eta_{\a,\b} \
\frac{\vartheta[{\a \atop \b }](\e) \vartheta[{{\a +3v_2} \atop \b }]}
{\vartheta[{1/2 \atop 1/2 }](\e) \eta^3} \right) \!\!\!
\prod_{i=1,2} \! \sin \pi 3 v_i
\frac{\vartheta[{\a +3v_i \atop \b }]}{\vartheta[{1/2 +3v_i \atop 
 1/2}]} W_3^{(2)} \nonumber
\ea
for the ${\cal N}=2$ sector $k=3$, where $v_2$($v_3$) is the volume (in string
units) of the second (third) compact torus.  
The corresponding Neumann-Dirichlet annulus amplitudes are
\ba
{}\!\!\!\!{1 \over l} {\cal A}_{95}^{(k)} \!\!&\!\!\!=\!\!&\!\!\!  {-iB \over 2
\pi^3}  {\rm tr} \left( Q \g_9^k \otimes \g_5^k  \!\!\!\!
\sum_{\a,\b=0,1/2} \!\!\!\!  \eta_{\a,\b} \!
\frac{\vartheta[{\a \atop \b }](\e)}{\vartheta[{1/2 \atop 1/2 }](\e)}
\right) \!
\frac{ \!\!\sin \pi k v_3 \vartheta[{\a +kv_3 \atop \b }]}{\vartheta[{1/2 +kv_3 \atop 
1/2 }]}
\prod_{i=1,2} \frac{\vartheta[{\a +kv_i \atop {\b + 1/2}}]}{\vartheta[{{1/2 + kv_i} \atop 
0 }]}  \  \label{a24}
\ea
for the sectors $k=1,2,4,5$ and
\ba
{}\!\!\!{1 \over l} {\cal A}_{95}^{(k)} \!\!&\!\!\!=\!\!&\!\!\! {B v_3 \over 8
\pi^3} {\rm tr} \left( Q \g_9^k \otimes \g_5^k
 \!\!\!\!
\sum_{\a,\b=0,1/2} \!\!\!\!  \eta_{\a,\b} \
\frac{\vartheta[{\a \atop \b }](\e) \vartheta[{\a +kv_3 \atop \b }]}
{\vartheta[{1/2 \atop 1/2 }](\e) \eta^3} \right) \!
\ \prod_{i=1,2} \  \frac{\vartheta[{\a +kv_i \atop {\b + 1/2}}]}{\vartheta[{{1/2 + kv_i} 
\atop 0 }]} \ W_3^{(2)}  \  \label{a25}
\ea
for $k=0,3$. The Neumann M{\"o}bius amplitudes read
\ba
{}\!\!\! {1 \over l} {\cal M}_9^{(k)} \!\!&\!\!\!=\!\!&\!\!  {8iB \over
\pi^3} {\rm tr} \left( Q \g_9^{2k} \!\!\!\!
\sum_{\a,\b=0,1/2}  \eta_{\a,\b} \
\frac{\vartheta[{\a \atop \b }] ({\e \over 2})}{\vartheta[{1/2 \atop 1/2
}] ({\e \over 2})} \right) 
\prod_{i=1}^3 \!\! \sin \pi k v_i
\  \frac{\vartheta[{\a +2kv_i \atop {\b + kv_i}}]}{\vartheta[{1/2 +2kv_i 
\atop {1/2 + kv_i}}]} \ , \label{a26}
\ea
for $k=1,5$,
\ba
{1 \over l} {\cal M}_9^{(k)} \!\!&\!\!\!=\!\!&\!\!\! -{4B v_2 \over \pi^3} 
{\rm tr} \left( Q \g_9^{2k}
 \!\!\!\!
\sum_{\a,\b=0,1/2} \!\!\!\!  \eta_{\a,\b}
\frac{\vartheta[{\a \atop \b }]({ \e \over 2}) \vartheta[{\a +2kv_2 \atop \b +kv_2}]}
{\vartheta[{1/2 \atop 1/2 }]({ \e \over 2}) \eta^3} \right) \!\!
\prod_{i=1,3} \!\! \sin \pi k v_i
\  \frac{\vartheta[{\a +2kv_i \atop {\b + kv_i}}]}{\vartheta[{1/2 +2kv_i \atop 
{1/2 + kv_i}}]} W_2^{(2),e} \nonumber
\ea
for $k=2,4$, where $W_2^{(2),e}$ denote the even winding sum along the second
compact complex coordinate and
\ba
{}\!\!\! {1 \over l} {\cal M}_9^{(3)} \!\!&\!\!\!=\!\!&\!\!\! -{4B v_3 \over
\pi^3} {\rm tr} \left( Q \g_9^{6}
  \!\!\!\!
\sum_{\a,\b=0,1/2} \!\!\!\!  \eta_{\a,\b}
\frac{\vartheta[{\a \atop \b }]({ \e \over 2}) \vartheta[{\a \atop \b +3v_2}]}
{\vartheta[{1/2 \atop 1/2 }]({ \e \over 2}) \eta^3} \right) \!
\prod_{i=1,2} \! \sin \pi 3 v_i
\frac{\vartheta[{\a \atop {\b + 3v_i}}]}{\vartheta[{1/2 \atop 
{1/2 + 3v_i}}]} W_3^{(2),e} \nonumber
\ea
for $k=3$.

%%%%%%%%%%%%%%%%%%%%%%%%%%%%%%%%%%%%%%%%%%%%%%%%%%%%%%%%%%%%%%%%%%%%%%%%%%%%%
%\newpage

%%%%%%%%%%%%%%%%%%%%%%%%%%%%%%%%%%%%%%%%%%%%%%%%%%%%%%%%%%%%%%%%%%%%%%%%%
\end{document}